\documentclass[showpacs,preprintnumbers,amsmath,amssymb,floatfix]{revtex4}

\usepackage[german, english]{babel}
\usepackage{graphicx}

\begin{document}

\title{Stable Phases of Boson Stars}

\author{
{\bf Burkhard Kleihaus, Jutta Kunz and Stefanie Schneider}
}
\affiliation{
{Institut f\"ur Physik, Universit\"at Oldenburg, 
D-26111 Oldenburg, Germany}
}
\date{\today}

\begin{abstract}
We analyze the physical properties of boson stars,
which possess counterparts in flat space-time, $Q$-balls.
Applying a stability analysis via catastrophe theory, we show that
the families of rotating and non-rotating boson stars 
exhibit two stable regions,
separated by an unstable region.
Analogous to the case of white dwarfs and neutron stars,
these two regions correspond to compact
stars of lower and higher density.
Moreover, the high density phase ends
when the black hole limit is approached.
Here another unstable phase is encountered,
exhibiting the typical spiralling phenomenon
close to the black hole limit.
When the interaction terms in the 
scalar field potential become negligible,
the properties of mini boson stars are recovered,
which possess only a single stable phase.
\end{abstract}

\maketitle



                  \section{Introduction}

After stars have consumed their nuclear fuel,
they end their lives as compact astrophysical objects,
and turn into white dwarfs, neutron stars (or their variants) or black holes,
depending on their initial mass.
When one investigates the equilibrium properties of
compact stellar configurations,
as discussed for instance in Shapiro and Teukolsky \cite{Shapiro:1983du},
one finds two stable phases for such compact stellar objects.
These stable phases correspond to the white dwarf phase
and the neutron star phase,
where equilibrium is achieved
by the electron and the neutron degeneracy pressure,
respectively.
The two stable phases are separated by an 
intermediate unstable phase.
Moreover, the stable neutron star phase is followed by 
an unstable phase, exhibiting a spiralling behavior
(when the mass is considered as a function of the radius).
As seen in Fig.~1, the stable neutron star phase ends
and the unstable phase sets in, when the black hole limit
is approached.

\begin{figure}[h!]
\centering
\includegraphics[width=70mm,angle=0,keepaspectratio]{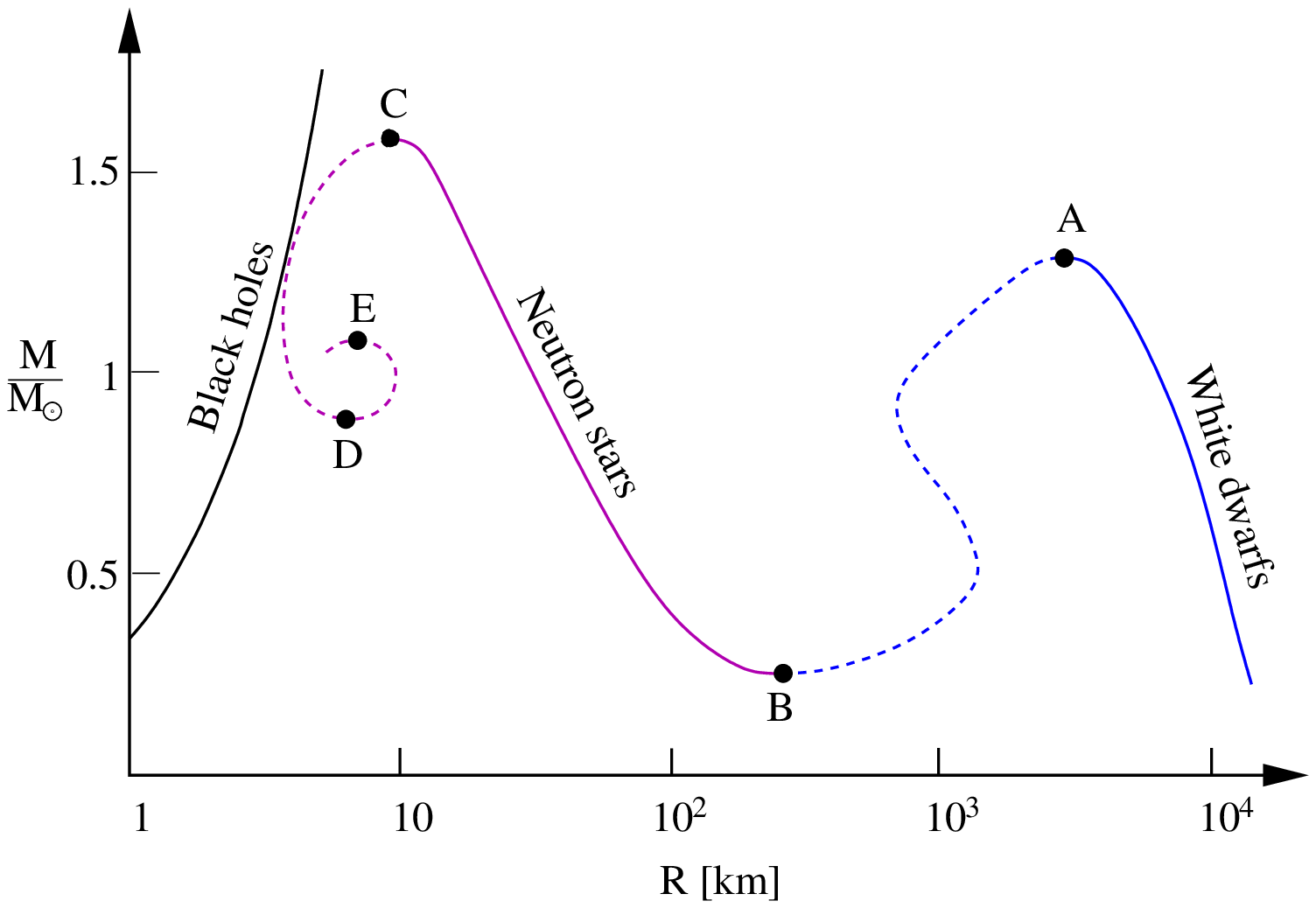}
\includegraphics[width=85mm,angle=0,keepaspectratio]{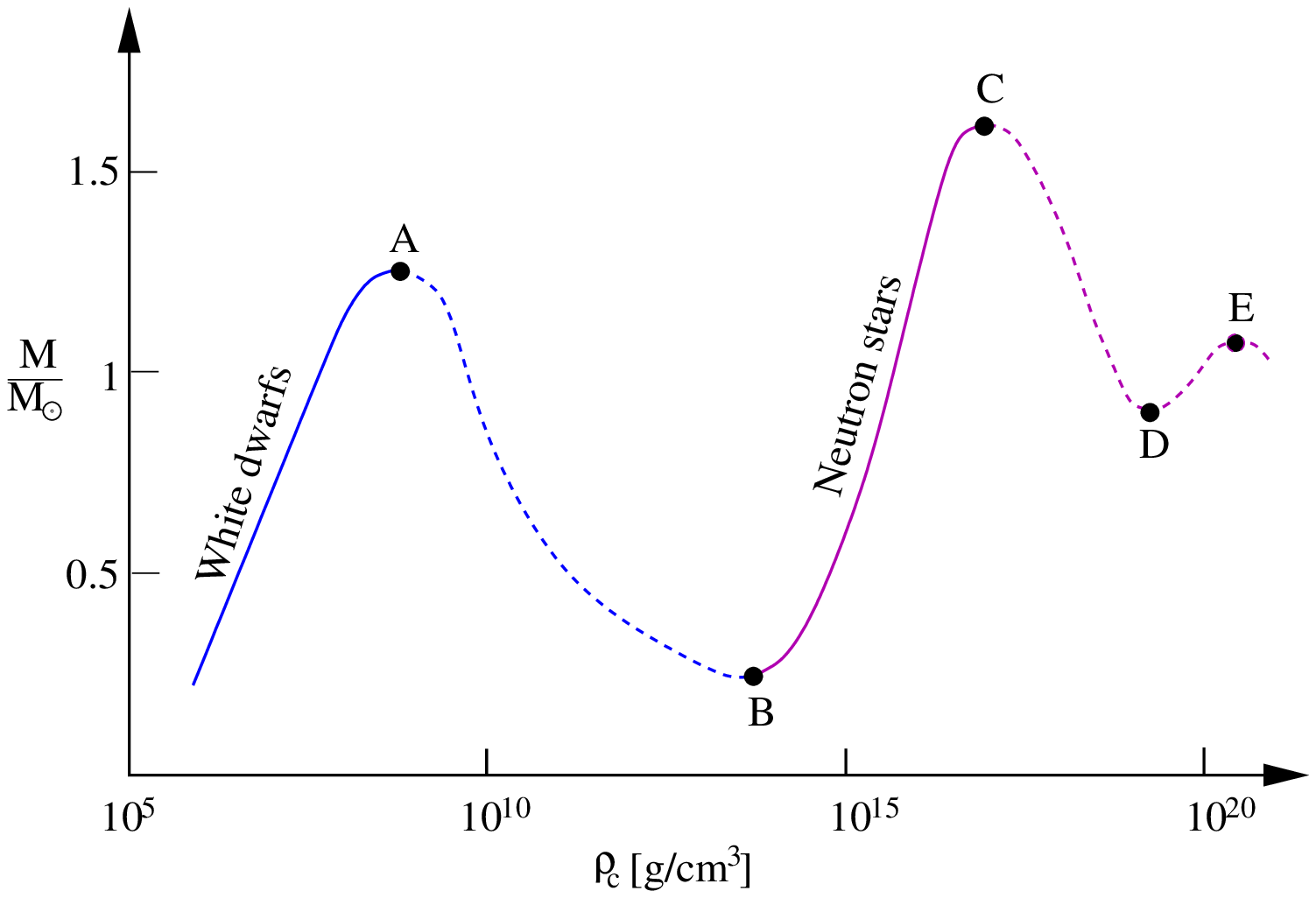}
\caption{Schematic diagram of the mass versus the radius (left)
and the mass versus the central density (right)
of compact stars after Shapiro and Teukolsky \cite{Shapiro:1983du}.}
\label{Shapiro}
\end{figure}

Representing hypothetical astrophysical objects,
boson stars have received considerable attention since 
being proposed by Feinblum and McKinley \cite{Feinblum:1968},
Kaup \cite{Kaup:1968zz},
and Bonazzola and Ruffini \cite{Ruffini:1969qy}.
Boson stars are obtained when a massive complex scalar field
is coupled to gravity.
The conserved Noether current associated with the global U(1)
symmetry is related to their particle number $Q$.
The physical properties of boson stars depend crucially on
the rest mass of the bosons and on the presence and type
of self-interaction of the bosons
(see e.g.~the review articles
\cite{Lee:1991ax,Jetzer:1991jr,Liddle:1993ha,Mielke:1997re,Schunck:2003kk}).

When only a mass term is present but no self-interaction
of the scalar field, the resulting family of mini-boson stars
possesses one stable branch of solutions,
which ends at the maximal mass configuration and is followed by an
unstable set of solutions with spiralling behaviour.
The maximal mass $M_{\rm max}$ is on the order of 
the Planck mass squared, divided by the boson mass:
$M_{\rm Pl}^2/m_{\rm B}$.
When a repulsive quartic self-interaction is included
the resulting family of boson stars
retains a single stable branch,
but the maximal mass $M_{\rm max}$ is now on the order of
$\sqrt{\lambda} M_{\rm Pl}^3/m_{\rm B}^2$,
where $\lambda$ is the self-coupling constant
\cite{Colpi:1986ye}.
Thus as shown by Colpi, Shapiro, and Wasserman
\cite{Colpi:1986ye} boson stars with much larger masses 
are obtained.

Friedberg, Lee and Pang \cite{Friedberg:1986tq}
considered boson stars with
a self-interaction of the scalar field
that allows for non-topological soliton solutions,
also called $Q$-balls, even in the absence of gravity.
They estimated that the maximal mass of such boson stars
is on the order of $M_{\rm Pl}^4/m_{\rm B}^3$
\cite{Lee:1991ax}.

Boson stars are obtained as stationary solutions 
of the coupled Einstein-scalar field equations,
when the scalar field has a harmonic time dependence.
The associated frequency $\omega_s$ is bounded from above
by the scalar mass $m_{\rm B}$,
while its lower bound depends on the details of the model.
Boson stars may rotate 
\cite{Schunck:1996wa,Ryan:1996nk,Yoshida:1997qf,Schunck:1999pm,Mielke:2000mh,Kleihaus:2005me,Kleihaus:2007vk}.
Since their total angular momentum is quantized in terms of their
particle number, $J=nQ$,
boson stars may change their angular momentum only in discrete steps
\cite{Schunck:1996wa}.

The stability of boson stars has been addressed from various points of view.
Whereas Lee and Pang \cite{Lee:1988av}
performed a linear stability analysis of boson stars
with respect to small oscillations,
Kusmartsev, Mielke and Schunck \cite{Kusmartsev:2008py,Kusmartsev:1992}
applied catastrophe theory to extract the stable 
branches of families of boson stars.
Catastrophe theory had been introduced by Rene Thom in the 1960's 
\cite{Thom:1975},
with further developments made by Zeeman \cite{Zeeman:1977},
Poston and Stewart \cite{Poston:1978,Stewart:1981},
Arnol'd \cite{Arnol'd:1985,Arnol'd:1992}, 
and many others.
Applications to solitons had been discussed by Kusmartsev 
\cite{Kusmartsev:1989nc}.

Here we analyze the properties and in particular the stability of
boson stars obtained with quartic and sextic 
self-interaction terms, 
as introduced by Friedberg, Lee and Pang 
in their non-topological soliton model
\cite{Friedberg:1986tq,Volkov:2002aj}.
Indeed, the existence of a flat space-time limit
is of profound importance for the properties of these boson stars,
as we will show below.
With respect to stability
we adapt the procedure of Tamaki and Sakai
\cite{Tamaki:2010zz,Tamaki:2011zza}.
They analyzed the stability of spherically symmetric Q-balls 
and boson stars via catastrophe theory 
under primarily mathematical points of view.
Here we extend this analysis to rotating boson stars.

In Section II we present the action, the equations of motion
and the definition of the global charges 
for non-rotating and rotating boson stars.
In Section III we analyze the physical properties
of the families of boson stars,
presenting first the results for the non-rotating case
and then for the rotating case.
We give our conclusions in Secion IV.
The Appendix addresses briefly the construction of the solutions.

\section{Model}

                      \subsection{Action}\label{c1s1}

We consider the action of a self-interacting complex scalar field 
$\Phi$ coupled to Einstein gravity
\begin{equation}
S=\int \left[ \frac{R}{16\pi G}
   -\frac{1}{2} g^{\mu\nu}\left( \Phi_{, \, \mu}^* \Phi_{, \, \nu} + \Phi _
{, \, \nu}^* \Phi _{, \, \mu} \right) - U( \left| \Phi \right|) 
 \right] \sqrt{-g} d^4x
\ , \label{action}
\end{equation}
where $R$ is the curvature scalar,
$G$ is Newton's constant,
the asterisk denotes complex conjugation,
\begin{equation}
\Phi_{,{\mu}}  = \frac{\partial \Phi}{ \partial x^{\mu}}
 \ ,
\end{equation}
and $U$ denotes the potential
\begin{equation}
U(|\Phi|) =  \lambda |\Phi|^2 \left( |\Phi|^4 -a |\Phi|^2 +b \right)
=\lambda \left( \phi^6- a\phi^4 + b \phi^2 \right) \ ,
\label{U} 
\end{equation}
with $|\Phi|=\phi$.
The potential is chosen such that nontopological soliton solutions
\cite{Friedberg:1986tq}, also referred to as $Q$-balls,
exist in the absence of gravity.
The potential has a minimum, $U(0)=0$, at $\Phi =0$
and a second minimum at some finite value of $|\Phi|$.
The boson mass is given by $m_{\rm B}=\sqrt{\lambda b}$.


Variation of the action with respect to the metric
leads to the Einstein equations
\begin{equation}
G_{\mu\nu}= R_{\mu\nu}-\frac{1}{2}g_{\mu\nu}R = \bar \kappa \, T_{\mu\nu}
\ , \label{Einstein}
\end{equation}
with 
$\bar \kappa = 8\pi G$
and stress-energy tensor $T_{\mu\nu}$
\begin{eqnarray}
T_{\mu \nu} &=& \phantom{-} g_{\mu \nu} L_M 
-2 \frac{ \partial L}{\partial g^{\mu\nu}}
 \\
&=&-g_{\mu\nu} \left[ \frac{1}{2} g^{\alpha\beta} 
\left( \Phi_{, \, \alpha}^*\Phi_{, \, \beta}+
\Phi_{, \, \beta}^*\Phi_{, \, \alpha} \right)+U(\phi)\right]
 + \left( \Phi_{, \, \mu}^*\Phi_{, \, \nu}
+\Phi_{, \, \nu}^*\Phi_{, \, \mu} \right
) \ .
\label{tmunu} \end{eqnarray}
Variation with respect to the scalar field
leads to the matter field equation,
\begin{equation}
\left(\Box+\frac{\partial U}{\partial\left|\Phi\right|^2}\right)\Phi=0 \ ,
\label{field_phi}
\end{equation}
where $\Box$ represents the covariant d'Alembert operator. 
Equations (\ref{Einstein}) and (\ref{field_phi}) represent
the set of the coupled Einstein--Klein--Gordon equations.

Also boson stars with two complex scalar fields
have been considered, leading to interesting
phenomena due to their interaction
\cite{Brihaye:2007tn,Brihaye:2008cg,Brihaye:2009yr}.
The influence of a negative cosmological constant
has been investigated in \cite{Astefanesei:2003qy}.

\subsection{Ansatz}\label{c2s1}

To obtain stationary axially symmetric solutions,
we impose on the space-time the presence of
two commuting Killing vector fields,
$\xi$ and $\eta$, where
\begin{equation}
\xi=\partial_t \ , \ \ \ \eta=\partial_{\varphi}
\   \label{xieta} \end{equation}
in a system of adapted coordinates $\{t, r, \theta, \varphi\}$.
In these coordinates the metric is independent of $t$ and $\varphi$,
and can be expressed in isotropic coordinates
in the Lewis--Papapetrou form 
\cite{Kleihaus:1997mn,Kleihaus:1997ws,Kleihaus:2000kg,Kleihaus:2002ee}
\begin{equation}
ds^2 = - f dt^2 
+ \frac{l}{f} ~ \left[ h \left( dr^2 + r^2 ~ d\theta^2 \right) 
  + r^2 ~ \sin^2 \theta ~ \left( d \varphi
- \frac{\omega}{r} ~ dt \right)^2 \right] \ . \label{ansatzg}
\end{equation}
The four metric functions $f$, $l$, $h$ and $\omega$
are functions of the variables $r$ and $\theta$ only.

The symmetry axis of the spacetime, where $\eta=0$, 
corresponds to the $z$-axis.
The elementary flatness condition 
\begin{equation}
\frac{X,_\mu X^{, \, \mu}}{4X} = 1 \ , \ \ \
X=\eta^\mu \eta_\mu \
\    \label{regcond} \end{equation}
then imposes on the symmetry axis the condition \cite{Kleihaus:1997mn}
\begin{equation}
h|_{\theta=0}=h|_{\theta=\pi}=1 \ .
\end{equation}

For the scalar field $\Phi$ we adopt the stationary ansatz \cite{Schunck:1996wa}
\begin{equation}
\Phi (t,r,\theta, \varphi) =  \phi (r, \theta) ~
 e^{ i\omega_s t +i n \varphi}  \label{ansatzp}
\end{equation}
where $\phi (r, \theta)$ is a real function,
and $\omega_s$ and $n$ are real constants.
Single-valuedness of the scalar field requires
\begin{equation}
\Phi(\varphi)=\Phi(2\pi + \varphi) \ , 
\end{equation}
thus the constant $n$ must be an integer,
i.e., $n \, = \, 0, \, \pm 1, \, \pm 2, \, \dots$.
We refer to $n$ as rotational quantum number,
since for $n \ne 0$ axially symmetric rotating boson stars arise, 
whereas for $n=0$ spherically symmetric non-rotating
boson stars are obtained.

Solutions with positive and negative parity satisfy, respectively,
\begin{eqnarray}
\phi (r, \pi-
\theta) & = & {\phantom{-}} \phi (r, \theta) \\
\phi (r, \pi-\theta) & = &           -   \phi (r, \theta) 
\ . \label{parity}
\end{eqnarray}

\subsection{Global charges}\label{c2s2}

The mass $M$ and the angular momentum $J$
of stationary asymptotically flat space-times
can be obtained from their respective Komar expressions \cite{Wald:1984},
\begin{equation}
{M} = 
 \frac{1}{{4\pi G}} \int_{\Sigma}
 R_{\mu\nu}n^\mu\xi^\nu dV
\ , \label{komarM1}
\end{equation}
and
\begin{equation}
{J} =  -
 \frac{1}{{8\pi G}} \int_{\Sigma}
 R_{\mu\nu}n^\mu\eta^\nu dV
\ . \label{komarJ1}
\end{equation}
Here $\Sigma$ denotes an asymptotically flat spacelike hypersurface,
$n^\mu$ is normal to $\Sigma$ with $n_\mu n^\mu = -1$,
$dV$ is the natural volume element on $\Sigma$,
$\xi$ denotes an asymptotically timelike Killing vector field
and $\eta$ an asymptotically spacelike Killing vector field
\cite{Wald:1984}.
The mass $M$ and the angular momentum $J$
can be read off 
directly from the asymptotic expansion of the metric functions $f$
and $\omega$, respectively,
\cite{Kleihaus:2000kg}
\begin{eqnarray}
M=\frac{1}{2G} \lim_{r \rightarrow \infty} r^2\partial_r \, f 
\ , \ \ \  J=\frac{1}{2G} \lim _{r \rightarrow \infty} r^2\omega \ .
\label{MJ2}
\end{eqnarray}

A conserved charge $Q$ is associated
with the complex scalar field $\Phi$,
since the Lagrange density is invariant under the global phase transformation
\begin{equation}
\displaystyle
\Phi \rightarrow \Phi e^{i\alpha} \ 
\end{equation}
leading to the conserved current
\begin{eqnarray}
j^{\mu} & = &  - i \left( \Phi^* \partial^{\mu} \Phi 
 - \Phi \partial^{\mu}\Phi ^* \right) \ , \ \ \
j^{\mu} _{\ ; \, \mu}  =  0 \ .
\end{eqnarray}
The conserved scalar charge $Q$ is obtained from the time-component
of the current,
\begin{eqnarray}
Q &=- & \int j^t \left| g \right|^{1/2} dr d\theta d\varphi 
\nonumber \\
 &=& 4 \pi \omega_s \int_0^{\infty}\int _0^{\pi} 
 |g| ^{1/2}   \frac{1}{f}  \left(  1 +
  \frac{n}{\omega_s}\frac{\omega}{r} \right) \phi^2 \,
dr \, d\theta \ . 
\label{Qc}
\end{eqnarray}

As first derived by Schunck and Mielke \cite{Schunck:1996wa},
one obtains a quantization relation for the angular momentum
in terms of the charge,
\begin{equation}
J=n \, Q \ . \label{JnQ}
\end{equation}
Thus a spherically symmetric boson star
has angular momentum $J =0$, because $n=0$.

\subsection{Units}

We choose for the potential $U(\phi)$,
Eq.~(\ref{U}), the following set of fixed parameters
\cite{Volkov:2002aj,Kleihaus:2005me,Kleihaus:2007vk}
\begin{equation}
\lambda =1, \quad a=2, \quad b=1.1=\mu_0^2 \ . 
\label{potpar} 
\end{equation}
The equations then depend only on the dimensionless
coupling constant $\kappa$
\begin{equation}
\kappa = 8 \pi G \left( \frac{m_{\rm B}}{\mu_0^2} \right)^2 \ .
\label{kappa} 
\end{equation}
Since $\kappa$ consists of a product of Newton's constant and the square
of the boson mass, we may interpret a change of its numerical
value in two ways: 
either as a change of the gravitational coupling
for a fixed boson mass, or as a change of the boson mass for
a fixed value of the gravitational coupling.

To obtain the respective dimensionful values of the physical properties,
$M^{\rm phys}$, $Q^{\rm phys}$, $J^{\rm phys}$,
we have to scale the numerically calculated values
for the mass $M^{\rm num}$, the charge $Q^{\rm num}$ 
and the angular momentum $J^{\rm num}$
appropriately.
Therefore we introduce the parameter $q_0$
\begin{equation}
 q_0=\left(\frac{m_{\rm Pl}}{m_{\rm B}}\right)^2\frac{1}{8\pi} \ ,
\label{q_0} 
\end{equation}
where $m_{\rm Pl}$ is the Planck mass and find
\begin{eqnarray}
 Q^{\rm phys} &=& \mu_0^2 \bar \kappa q_0 Q^{\rm num} \ , \nonumber \\
 M^{\rm phys} &=& \mu_0 \bar \kappa q_0 m_{\rm B} M^{\rm num} \ , \nonumber \\
 J^{\rm phys} &=& n\hbar Q^{\rm num} \ .
\end{eqnarray}
The length scale is set by
$$\sigma = \frac{\mu_0}{m_{\rm B}} 
\approx \frac{\mu_0}{m_{\rm B}[{\rm GeV}]}\times 0.2 \times 10^{-15} {\rm m}$$
Two examples for the scales of the physical properties 
are exhibited in Table 1.

\begin{center}
\begin{table}[h!]
\begin{tabular}{c|c|c|c}	    
 $m_{\rm B}$ & $q_0$ &  $q_0 m_{\rm B}$ & $ \sigma $\\[5pt]
\hline   
               ${\rm 1 GeV/c^2 }$  &
               ${\displaystyle \approx 10^{36}}$ &		    
               ${\displaystyle \approx 10^{36} {\rm GeV/c^2} }$ & 
               $  \approx 0.2 \mu_0 {\rm fm} $\\[3pt]
\hline
               $10^{-19}{\rm GeV/c^2} $  &
               ${\displaystyle \approx 4\times 10^{74}}$ &		    
               ${\displaystyle \approx M_{\rm sun}}$ & 
               $ \approx 0.2 \mu_0 10^{4} {\rm m}$ 
\end{tabular}
\caption{Scale factors for the physical properties}

\end{table}
\end{center}

\section{Boson star properties}


\subsection{Spherically symmetric boson stars}

\subsubsection{Equilibrium space}

Let us first
consider the families of fundamental spherically symmetric boson stars
($n = 0$) 
as obtained in \cite{Kleihaus:2005me,Kleihaus:2007vk}.
(In fundamental boson stars the scalar field is a monotonically 
decreasing function,
whereas in radially excited boson stars, it possesses nodes.)
For a fixed value of the dimensionless coupling constant 
$\kappa$, Eq.~(\ref{kappa}),
a family of stationary solutions exists in the frequency range
$\omega_0(\kappa) \le \omega_s \le \omega_{\rm max}$.
The minimal frequency $\omega_0(\kappa)$ depends on $\kappa$
and increases with $\kappa$,
tending to finite limits as 
$\kappa \to 0$ and $\kappa \to \infty$ \cite{Kleihaus:2005me}.
The maximal frequency $\omega_{\rm max}$ is always given by
the boson mass.

\begin{figure}[h!]
\begin{center}
\includegraphics[width=0.65\textwidth, angle =0]{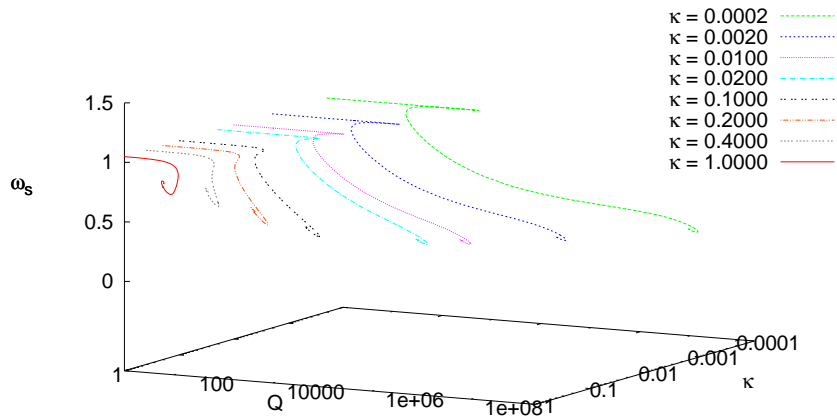}
\end{center}
\vspace{-1.0cm}
\caption{\small
{\it Equilibrium space} $\mathcal{M} = \{\omega_s, Q, \kappa \}$ 
for fundamental boson stars 
in the range $0.0002 \le \kappa \le 1$.}
\label{WhitFundBSohne0}
\end{figure}

All families of solutions together form the 
{\it equilibrium space}, which we denote by
$\mathcal{M} = \{\omega_s, \kappa, Q \}$.
To illustrate the {\it equilibrium space} 
we have exhibited several families of solutions
in Fig.~\ref{WhitFundBSohne0}.
For each family (with fixed $\kappa$)
the frequency $\omega_s$ is shown versus the
particle number $Q$.
The coupling constant $\kappa$ assumes values
in the range $0.0002 \le \kappa \le 1$ in the figure.
The equilibrium space may then be pictured
as the surface obtained when $\kappa$ varies continuously
from zero to infinity.

The families of solutions have two important features in common.
They all start from particle number $Q=0$ at the upper limit
of the frequency, and they all form spirals
towards their lower frequency limit.
Inbetween, however, a qualitative change of the curves is observed,
as the value of $\kappa$ increases.
The salient maximum value of $Q$ 
in the small frequency range 
as well as the finite (relative) minimum value of $Q$, 
which are both present for small values of the coupling constant $\kappa$,
become less pronounced with increasing $\kappa$,
until they merge and disappear altogether.

We may also consider the set of equilibrium solutions
with respect to another set of parameters.
Instead of the frequency $\omega_s$ we may consider the
finite value $\phi_0=\phi(0)$ of the scalar field at the origin.
(We recall, that in these fundamental boson star solutions 
the scalar field decreases monotically from the origin
to zero at infinity.)
Moreover, instead of the particle number $Q$
we may consider the mass $M$ of the solutions.

\begin{figure}[h!]
\begin{center}
\includegraphics[width=0.65\textwidth, angle =0]{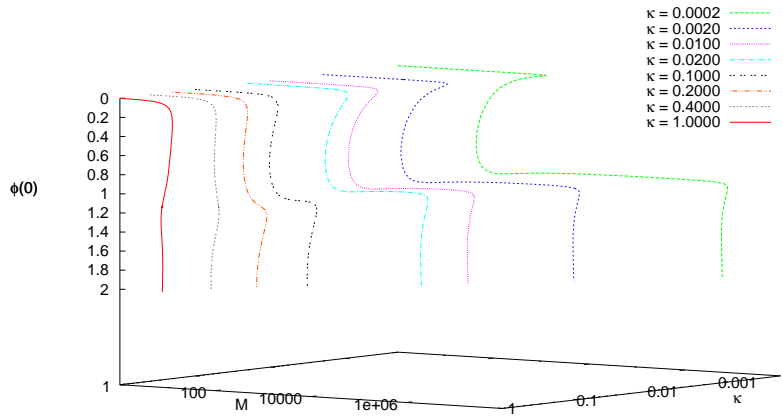}
\vspace{-1.0cm}
\caption{\small 
{\it Equilibrium space} $ \mathcal{N} = \{ \phi_0, M, \kappa \} $
for fundamental boson stars in the range $0.0002 \le \kappa \le 1$. }
\label{MueberfWhitney}
\end{center}
\end{figure}

We exhibit the same families of solutions as above
for this choice of variables in Fig.~\ref{MueberfWhitney}.
Here for each family of solutions (with fixed $\kappa$)
the value of the scalar field $\phi_0$ is shown versus the
mass $M$.
We note, however, that the range of $\phi_0$ has been truncated
in the figure, being limited to the range $0 \le  \phi_0 \le 2$.
Inside the spiral $\phi_0$ certainly assumes larger values,
and indeed increases monotonically as the mass exhibits damped oscillations
\cite{Kleihaus:2005me}.
The equilibrium space $ \mathcal{N} = \{ \phi_0, M, \kappa \} $
may then again be pictured
as the surface obtained when $\kappa$ varies continuously
from zero to infinity.

\subsubsection{Binding energy and cusp structure}

\begin{figure}[h!]
\begin{center}
\includegraphics[width=0.45\textwidth, angle =0]{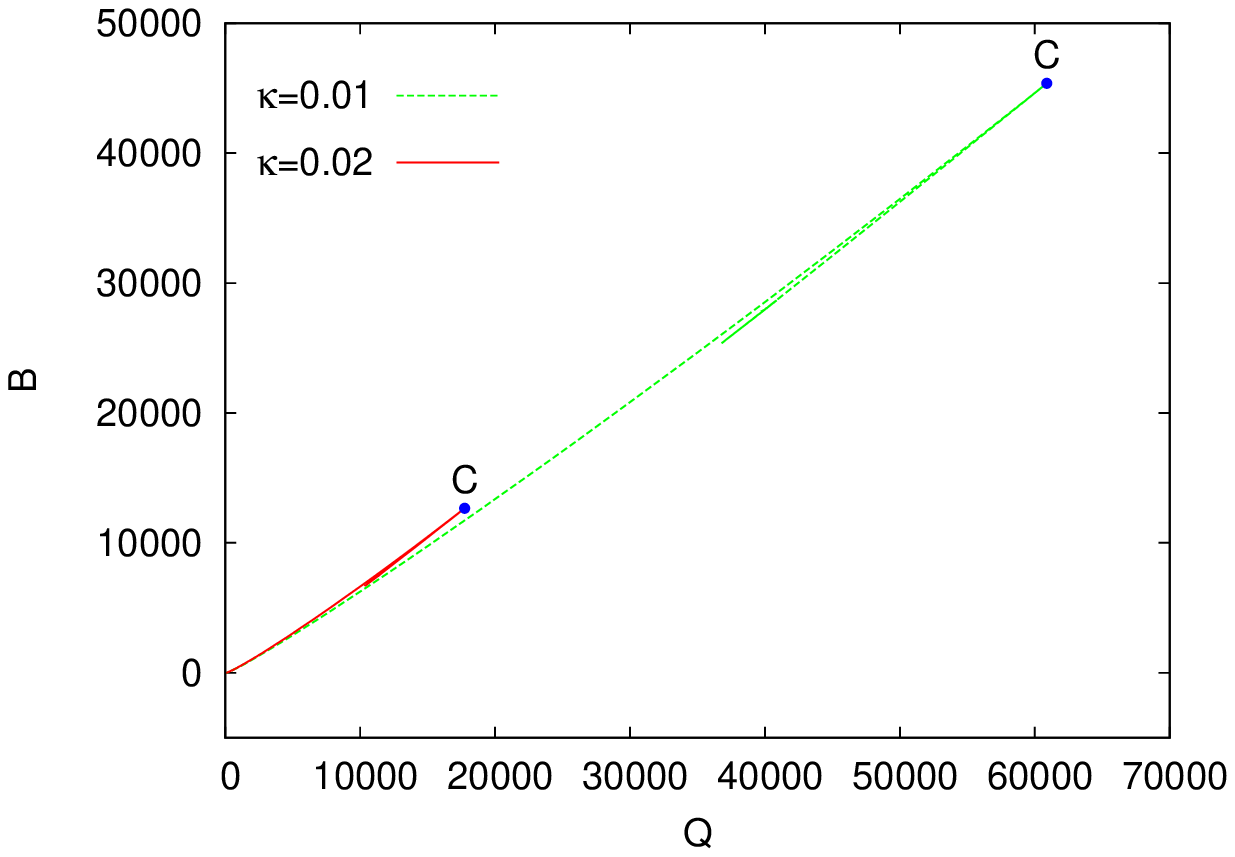}
\includegraphics[width=0.45\textwidth, angle =0]{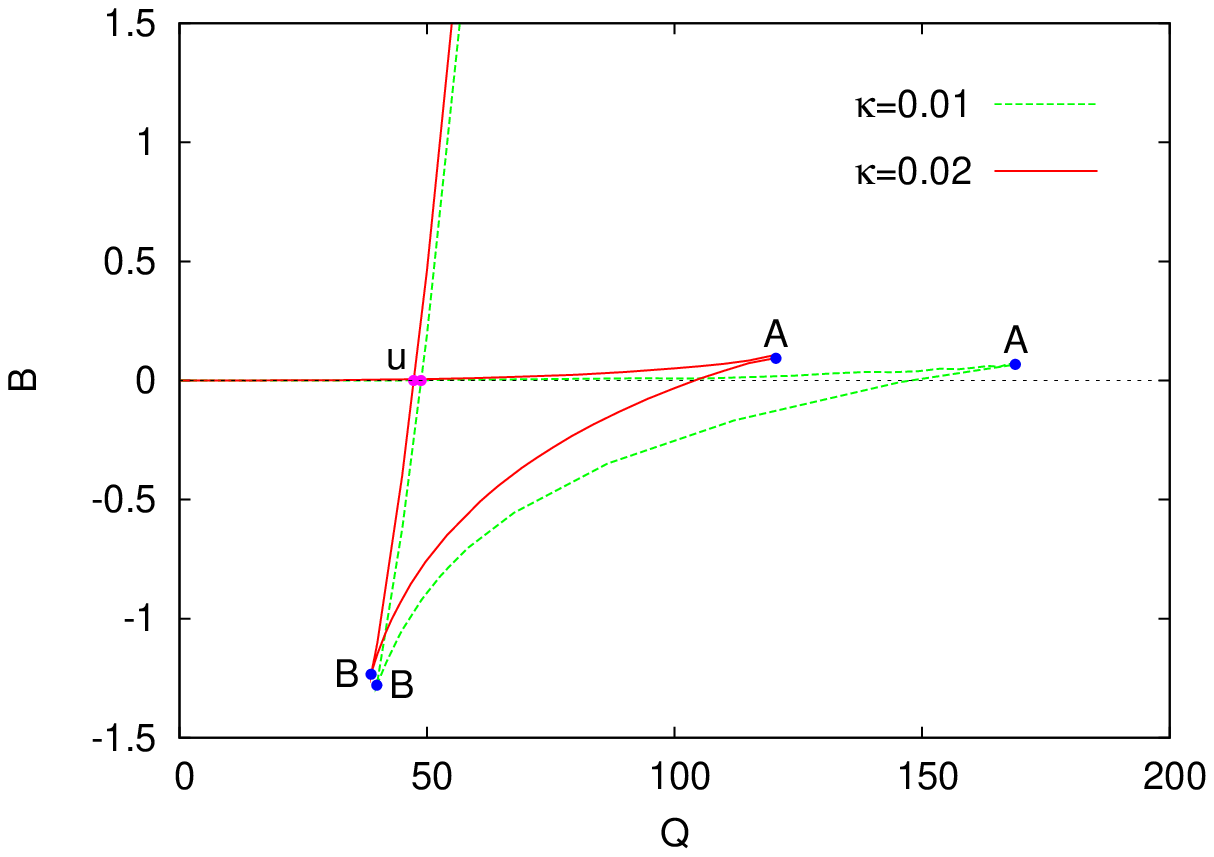}
\caption{\small
Binding energy $B=M-m_{\rm B}Q$ 
for two families of solutions, corresponding to
$\kappa=0.01$ and $\kappa=0.02$, respectively.
The right figure zooms into the high frequency range.}
\label{binding}
\end{center}
\end{figure}

To get a better physical understanding of these boson stars
that form the equilibrium space, let us next address
their binding energy $B=m_{\rm B}Q-M$.
We exhibit the binding energy $B$ in Fig.~\ref{binding}
for two families of solutions, corresponding to
typical examples in the lower $\kappa$ range,
namely $\kappa=0.01$ and $\kappa=0.02$.
While at first glance all solutions seem to be bound,
with their binding energy increasing almost linearly with
their particle number,
a closer look at the solutions in the high frequency range
($\omega_s$ close to $m_B$)
reveals a bifurcation structure
involving two cusps.

Clearly, a first branch of bound boson star solutions 
resides between the vacuum $M=Q=0$, 
and a local maximum of the mass and charge,
$M_A$, $Q_A$, at the first cusp $A$.
A second branch, where the mass and charge decrease monotonically,
connects the first cusp $A$ with the second cusp $B$,
where the mass and charge have a local minimum,
$M_B$, $Q_B$.
Most of this branch resides in the unbound region.
The third branch emerges from the second cusp $B$
and extends up to the third cusp $C$,
where the mass and charge reach their global maximum,
$M_C$, $Q_C$.
Along this third branch the mass and charge increase monotonically.
The solutions are unbound only in the vicinity of the second cusp $B$,
since the binding energy crosses the zero already
at the point ${u}$, located closeby.
At the third cusp $C$ also the binding energy is maximal.

\begin{figure}[h!]
\begin{center}
\includegraphics[width=0.4\textwidth, angle =0]{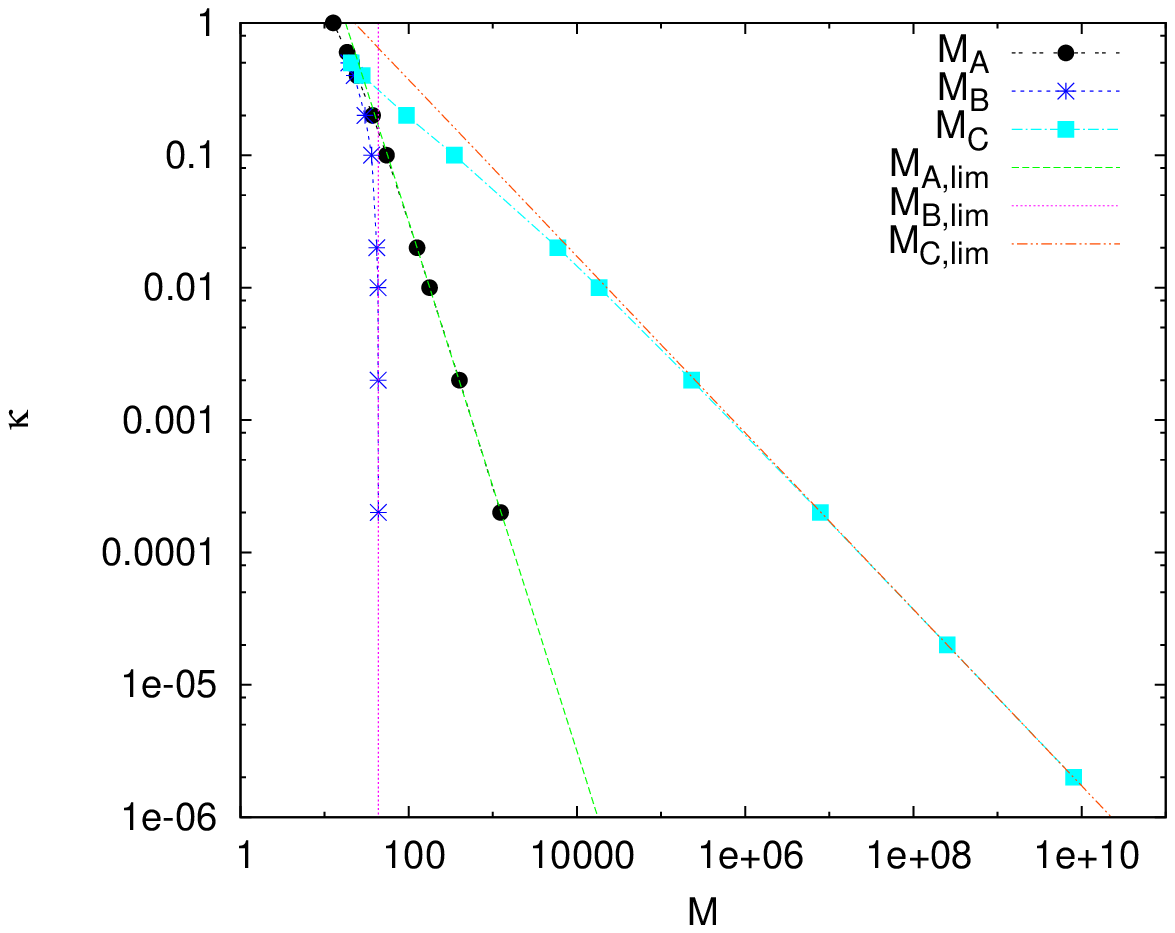}
\includegraphics[width=0.4\textwidth, angle =0]{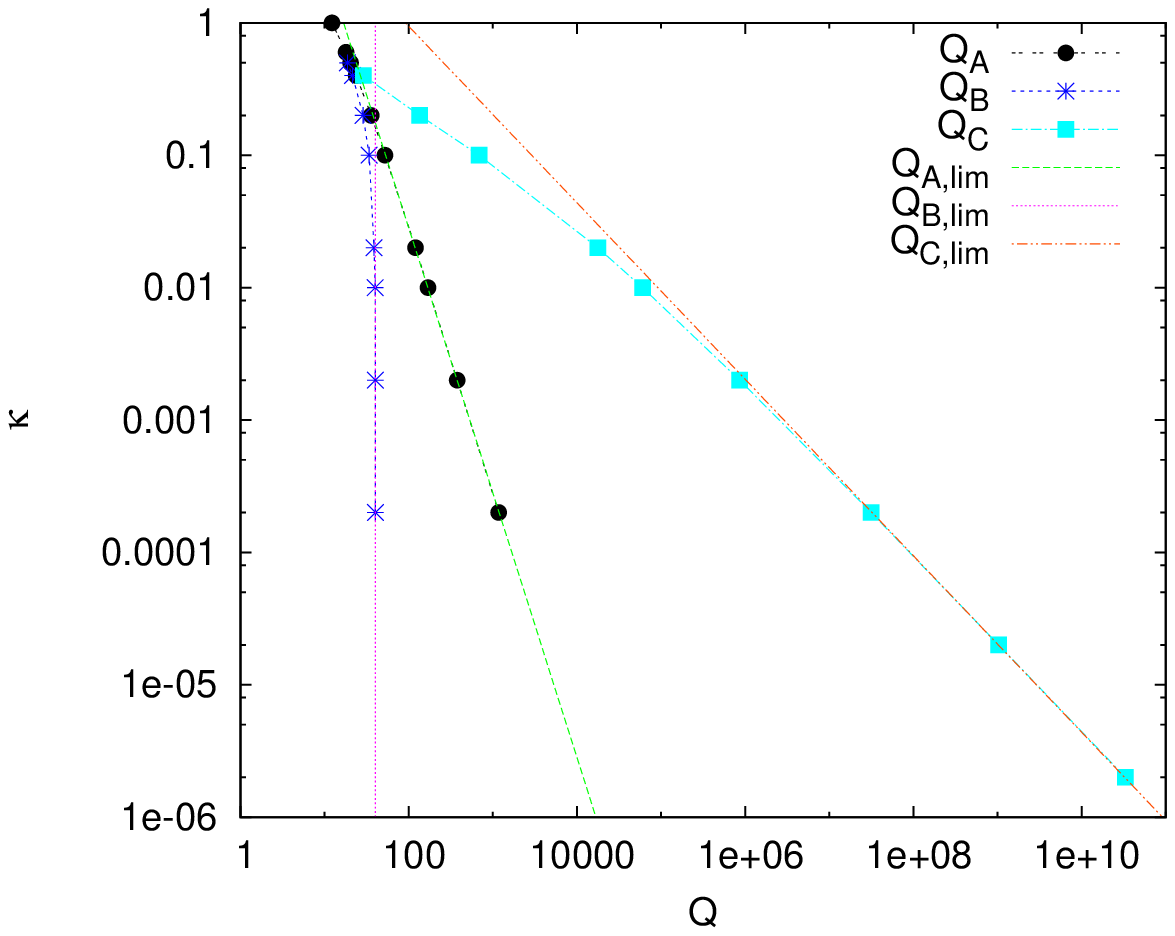}
\caption{\small
Coupling constant $\kappa$
versus the mass $M$ (left) and
versus the particle number $Q$ (right)
for the cusps $A$, $B$ and $C$,
together with functions approximating their behavior
in the limit $\kappa \to 0$.
Note, that for the larger values of $\kappa$
the cusps $B$ and $C$ are no longer present.}
\label{cuspsABC}
\end{center}
\end{figure}

We exhibit in Fig.~\ref{cuspsABC} the physical characteristics
of these 3 cusps, $A$, $B$ and $C$.
As $\kappa \to 0$, the values of the mass $M_A$ and the charge $Q_A$ at
the cusp $A$ increase, tending to infinity with the power $\kappa^{-1/2}$,
as illustrated in Fig.~\ref{cuspsABC} by the curves $M_{A,\rm lim}$
and $Q_{A,\rm lim}$.
The values of the mass $M_B$ and the charge $Q_B$ at the cusp $B$
show a very different behavior. They tend to constant values
in the limit $\kappa \to 0$, corresponding to the
values of the mass and the charge of the unique minimum
of the $Q$-ball solutions of flat space-time.
This is also illustrated in Fig.~\ref{cuspsABC}.
At the cusp $C$ the values of the mass $M_C$ and the charge $Q_C$
increase again without bound as $\kappa \to 0$.
As extrapolated previously \cite{Kleihaus:2005me},
the limiting behaviour is of the form
$M_C \sim \kappa^{-3/2}$, $Q_C \sim \kappa^{-3/2}$,
which is also shown in the figure.

Turning to the larger values of $\kappa$,
we note that the two extrema $B$ and $C$, and thus the two cusps
$B$ and $C$, merge and disappear
at a critical value $\kappa_{\rm cr}$.
Thus beyond $\kappa_{\rm cr}$, from the three cusps $A$, $B$ and $C$
only the cusp $A$ is left.
The large value of $\kappa$ means from a physical point of view,
that the higher order terms in the potential become relatively
less important, and the solutions tend to the well-known
mini-boson star solutions.
Indeed, for $\kappa \to \infty$ the mini-boson star solutions
are recovered after a rescaling \cite{Kleihaus:2005me}.

To complete the discussion of the branches of the families of 
boson star solutions, we still need to address their behavior
beyond the cusp $C$ 
(or for large $\kappa$ beyond the cusp $A$).
As seen in Fig.~\ref{WhitFundBSohne0}
the branches form a spiral when considered in terms of
the frequency $\omega_s$ and the charge $Q$ 
(or the mass $M$).
Alternatively, when considered in terms of the
value of the scalar field at the origin and the mass $M$
(or the charge $Q$)
as in Fig.~\ref{MueberfWhitney},
a damped oscillation is seen.
Thus beyond $C$ 
(or for large $\kappa$ beyond $A$)
the mass and the charge reach 
a minimum $D$, then another maximum $E$,
another minimum $F$, etc.,
converging towards limiting values
$M_{\rm lim}$, $Q_{\rm lim}$ \cite{Kleihaus:2005me}.
When the mass is considered as a function of the charge,
finally, this behavior translates into an intricate
cusp structure \cite{Friedberg:1986tq}.

\subsubsection{Size and black hole limit}

\begin{figure}[h!]
\begin{center}
\includegraphics[width=0.45\textwidth, angle =0]{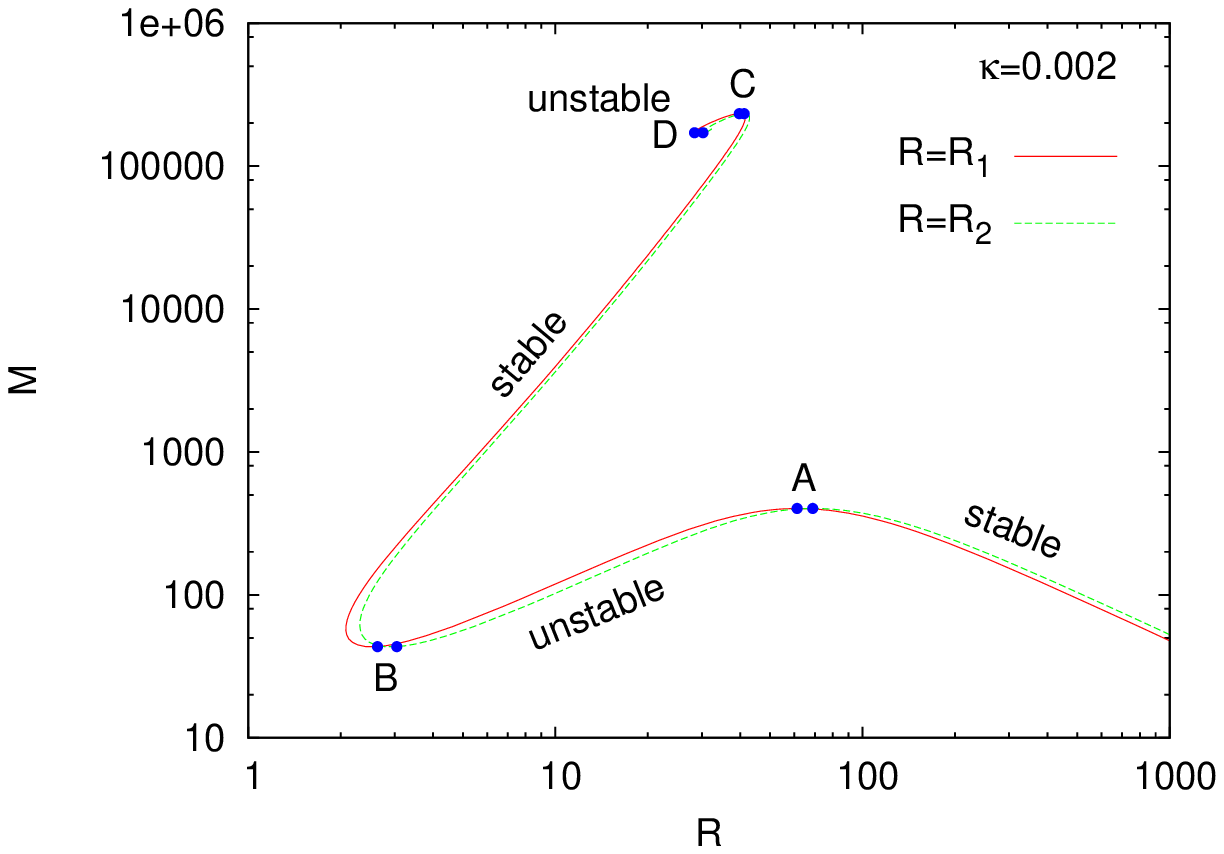}
\includegraphics[width=0.45\textwidth, angle =0]{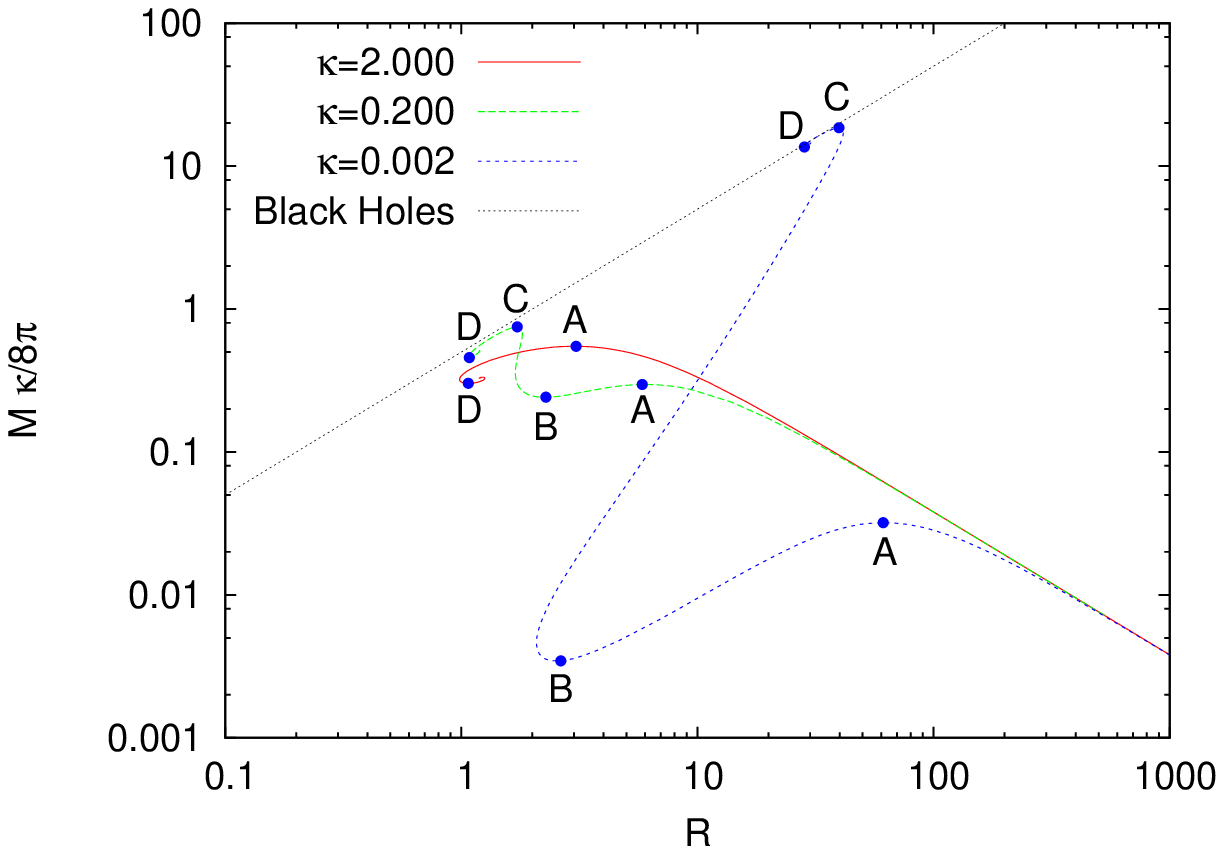}
\caption{\small
A comparison of
the two definitions Eq.~(\ref{radius}) for the boson star radius 
for the family of solutions with $\kappa=0.002$.
The cusps $A$ through $D$ are also marked (left).
The mass versus the boson star radius $R_1$ for the families of solutions
with $\kappa=1.0$, $\kappa=0.2$ and $\kappa=0.002$ together
with the corresponding Schwarzschild black hole curve (right).
}
\label{fig_radius}
\end{center}
\end{figure}

There is no unique definition for the radius of a boson star,
and many proposals have been discussed in the literature
(see e.g.~\cite{Schunck:2003kk}).
Here we have chosen 
the definitions 
\begin{equation}
R_1= \frac{\int j^t \left| g \right|^{1/2} r dr }
     {\int j^t \left| g \right|^{1/2} dr } \ , 
\quad\quad
R_2^2= \frac{\int j^t \left| g \right|^{1/2} r^2 dr }
     {\int j^t \left| g \right|^{1/2} dr } \ 
\label{radius}
\end{equation}
for the spherically symmetric boson stars,
where the radial coordinate is not the isotropic coordinate
introduced in Section II, but a Schwarzschild-like coordinate,
which has an invariant circumferential meaning.
Fig.~\ref{fig_radius} (left) shows, that both definitions
give rather similar results for the size of the boson stars.
From the radius and thus the size of the boson stars
together with their mass we obtain an estimate of their density.

Let us now discuss the relation between the mass and the size
of the boson stars, focussing on the smaller values of $\kappa$
(away from the mini-boson star limit).
As seen in Fig.~\ref{fig_radius},
the mass increases along the first branch,
while the radius decreases.
This is the behavior known from compact objects
such as white dwarfs and neutron stars.
The first branch ends at the cusp $A$.
Beyond $A$ the mass decreases until the
second cusp $B$ is reached, while the radius
continues to decrease along this second branch.
Beyond $B$ the mass increases again,
and soon rises steeply along this third branch.
But the radius (soon) starts to increase as well, 
and then rises further along this branch, 
almost up to the cusp $C$.
Thus while the boson stars are already compact objects on the first branch,
they become even much more compact objects along the third branch.

This behavior is very reminiscent of  
the phases of compact fermionic stars, where the lower density
stars represent white dwarfs, while the higher density stars
represent neutron stars or their variants.
The neutron stars seize to exist, when the black hole limit is reached.
Close to this limit, the solutions exhibit a spiralling behavior
as seen in the schematic drawing in Fig.~\ref{Shapiro}.
And indeed, when we include the black hole limit,
given by the Schwarzschild relation $2 G M=R$,
in the figure for the boson stars, Fig.~\ref{fig_radius} (right),
we observe that the spiral is precisely formed,
when the family of boson star solutions approaches 
the black hole limit.

\begin{figure}[h!]
\begin{center}
\includegraphics[width=0.45\textwidth, angle =0]{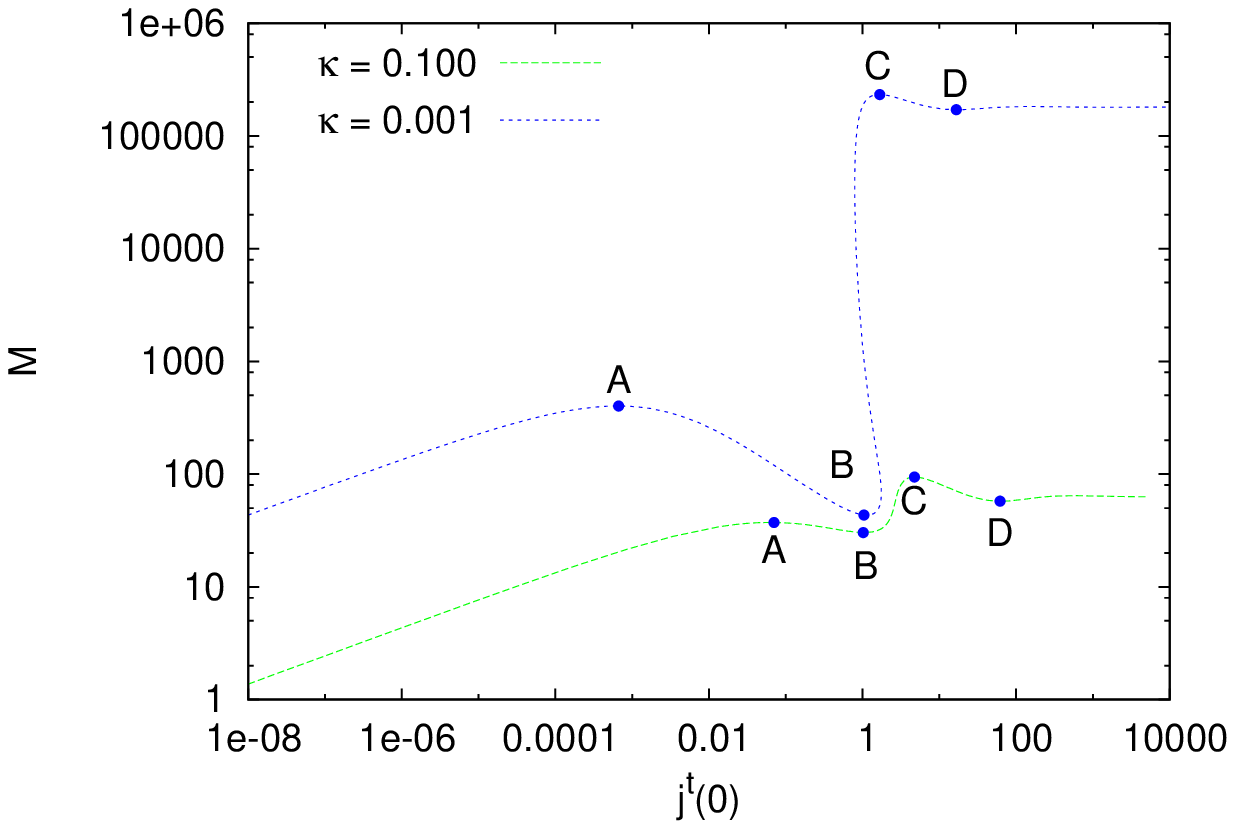}
\includegraphics[width=0.45\textwidth, angle =0]{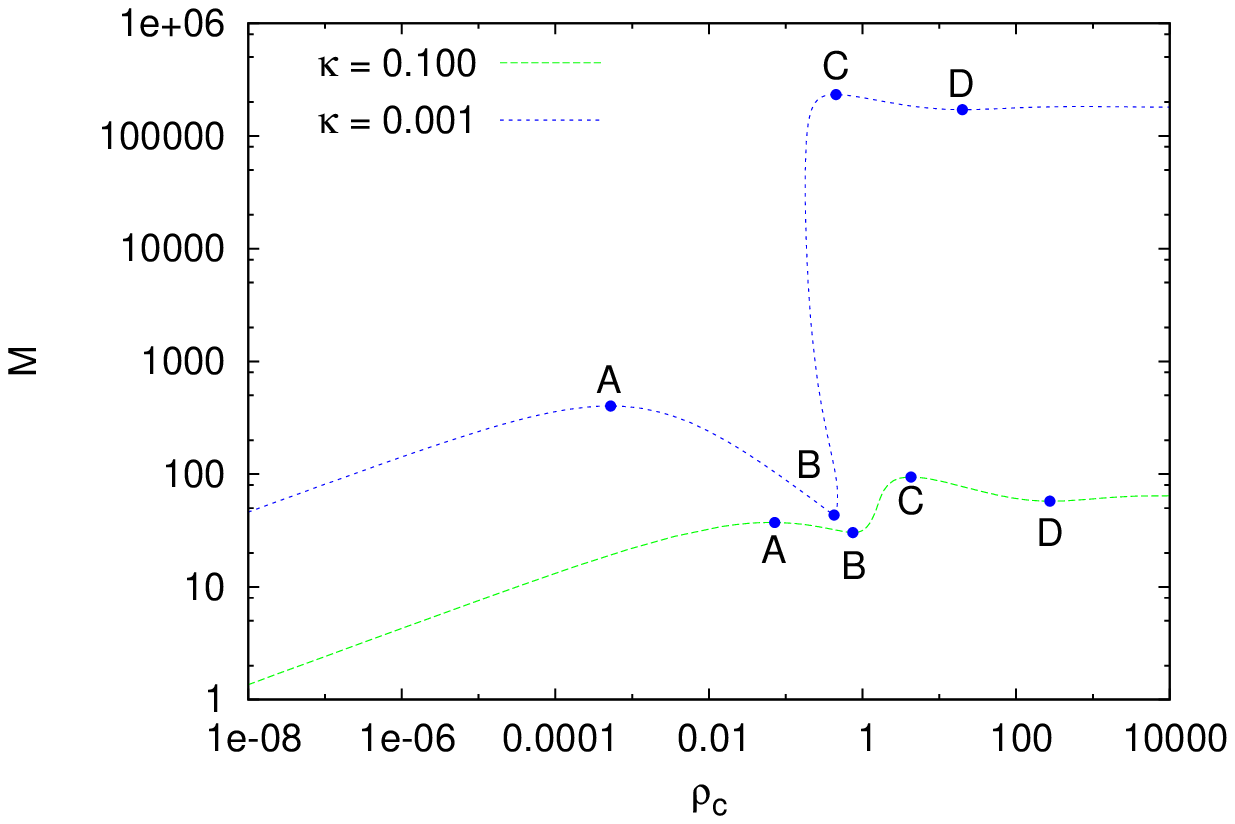}
\caption{\small
The mass versus the central particle number density
$j^t(0)$ (left)
and the central energy density $\rho_c(0)$ (right)
of the families of boson star solutions
with $\kappa=0.001$ and $\kappa=0.1$.}
\label{fig_radius_dens}
\end{center}
\end{figure}

Finally, in Fig.~\ref{fig_radius_dens} we exhibit
the mass of the boson stars versus the 
central particle number density $j^t(0)$ (left)
and versus the central energy density $\rho_c(0)$ (right).
Clearly, the first branch has much lower central density
than the third branch, just like 
the white dwarf branch and the neutron star branch
of the compact fermionic stars
\cite{Shapiro:1983du}.

\subsubsection{Stability analysis}

Let us now address the stability of these families of 
spherically symmetric boson stars.
The above observed analogy between the branches
of compact stars and the branches of boson stars, 
suggests to begin by recalling the stability properties of the
compact stars as indicated in Fig.~\ref{Shapiro},
following Shapiro and Teukolsky \cite{Shapiro:1983du}.

The white dwarf branch ending at the cusp $A$ is stable.
It is followed by an unstable branch, ending at the cusp $B$.
The neutron star branch from $B$ to $C$ is again a stable branch.
The spiral beyond $C$ is unstable.
In particular, one mode becomes unstable at $A$,
and turns stable again at $B$. 
At $C$ the mode becomes unstable again,
and at each following extremal point in the spiral
another mode turns unstable.
Thus there are two physically relevant stable branches,
the lower density white dwarf branch and the higher density
neutron star branch. The latter ends close to the black hole limit.

For mini-boson stars a mode analysis has been performed
by Lee and Pang \cite{Lee:1988av}
to determine the stability of the solutions.
Mini-boson stars possess only a single stable branch
followed by the spiral. As in the case of the compact fermionic stars,
at each following extremal point in the spiral
another mode of these boson stars turns unstable.

On the other hand, the stability of boson stars has been analyzed
by envoking the arguments of catastrophe theory
\cite{Kusmartsev:2008py,Tamaki:2010zz,Tamaki:2011zza}.
We briefly recall the procedure employed by Tamaki and Sakai
\cite{Tamaki:2010zz,Tamaki:2011zza},
applying it to the above family of boson star solutions.

An essential point in utilizing catastrophe theory 
is to select an appropriate set of 
\textit{behavior variable(s)} and \textit{control parameter(s)}. 
A behavior variable should be a quantity
that describes the behavior of the system uniquely 
when the control parameters change their values.
Following Tamaki and Sakai \cite{Tamaki:2010zz,Tamaki:2011zza},
we choose the coupling constant $\kappa$ and the charge $Q$ 
as the two control parameters,
and we choose the frequency $\omega_s$ as the single behavior variable.
(In the relevant range of solutions up to the cusp $C$, 
the variable $\omega_s$ is indeed unique, but not beyond.
In contrast, the variable $\phi_0$ is unique in the spiral
beyond the cusp $C$, but is not necessarily unique in the vicinity of
$C$ right before the cusp.)

To analyze the stability of these boson stars, we start from the 
\textit{equilibrium space} $\mathcal{M} = \{\omega_s, \kappa, Q \}$,
exhibited in Fig.~\ref{WhitFundBSohne0}.
According to catastrophe theory,
the stability changes only at the turning points,
where $\frac{\partial Q}{\partial \omega_s} = 0$
(for fixed values of $\kappa$). Therefore
\begin{itemize}
 \item we determine the turning points, where stability changes,
 denoting them by $A$, $B$, $C$, $\dots$;
 \item we plot the values of the turning points $A$, $B$, $C$, $\dots$
  versus $\kappa$ to obtain the 
  \textit{control space} $\mathcal{C} = \{\kappa, Q \}$,
  to read off the regions of stability, respectively, instability.
 \end{itemize}

We list the turning points in Table \ref{fig10QN}
together with the extrapolated values of the limiting solutions $Q_{lim}$ 
at the centers of the spirals \cite{Kleihaus:2005me}.

\begin{table}[h!]
\begin{center}
\begin{tabular}{|c|c|c|c|c|c|c||c|}
\hline
$\kappa$ & $Q_{A}$ & $Q_{B}$ & $Q_{C}$ & $Q_{D}$ & $Q_{E}$ & $Q_{F}$ & $Q_{lim}$   \\
\hline
\hline
$0.0002$ & $1174.78$ & $40.2495$ & $3.117 \cdot 10^{7}$ & $2.040 \cdot 10^{7}$ & $2.220 \cdot 10^{7}$ & $2.181 \cdot 10^{7}$ & $2.181 \cdot 10^{7}$ \\
$0.0020$ & $377.455$ & $40.0820$ & $864746$             & $553687$             & $605046$             & $593437$             & $5.954 \cdot 10^{5}$ \\
$0.0100$ & $169.834$ & $39.9989$ & $60416.2$            & $36845.6$            & $40849.6$            & $39949.6$            & $4.011 \cdot 10^{4}$ \\
$0.0200$ & $120.476$ & $38.7401$ & $17702.5$            & $10350.7$            & $11593.1$            & $11307.0$            & $1.136 \cdot 10^{4}$ \\  
$0.1000$ & $52.22$   & $33.91$   & $684.741$            & $331.677$            & $391.874$            & $378.274$            & $380.4600$           \\
$0.2000$ & $35.763$  & $28.668$  & $134.72$             & $61.19$              & $72.96$              & $70.463$             & $70.8330$            \\
$0.4000$ & $23.66$   & $21.47$   & $28.89$              & $15.85$              & $17.85$              & $17.46$              & $17.5180$            \\ \hline
$1.0000$ & $12.22$   & $-$       & $-$                  & $5.83$               & $6.31$               & $6.29$               & $6.2114$             \\

 \hline
\end{tabular}
\end{center}
\caption{ Values of the charge at the turning points $A$, $B$, $C$, $\dots$,
and of limiting solution $Q_{lim}$ at the center of the spiral 
for the fundamental boson stars for several values 
of the coupling constant $\kappa$. 
As indicated by the line, the cusps $B$ and $C$ have 
merged and disappeared between $\kappa=0.4$ and $\kappa=1$.
}
\label{fig10QN}
\end{table}

Each point of the eqilibrium space 
$\mathcal{M}= \{\omega_s, \kappa, Q \}$ 
represents a boson star configuration.
The set of turning points partitions this phase portrait into subareas.
In catastrophe theory,
``stability'' means stability with respect to local perturbations.
According to catastrophe theory,
passing a turning point means changing the stability 
of the boson star configurations.
The solutions between two turning points then form \textit{branches},
where all the configurations 
possess the same kind of stability (or instability).
Thus there are \textit{S}- and \textit{U}-branches 
representing stable and unstable configurations
\cite{Seidel:1990jh,Balakrishna:1997ej}.

This reveals the strength of catastrophe theory: 
stability changes exclusively when passing a turning point,
while all configurations of the considered system between two turning points 
possess the same kind of stability (or instability). 
Thus, it is sufficient to consider only a single configuration of a branch 
to know the stability of all the other configurations of the same branch
with respect to local perturbations.

For boson star solutions we see a cusp-catastrophe 
because (leaving the spiral aside)
their equilibrium space $\mathcal{M}$ 
shows the characteristics of a Whitney-surface 
\cite{Whitney:1955,Kusmartsev:1989nc}.
Therefore we conclude that the branch from the vacuum solution 
to the first turning point $Q_{A}$ is a stable one 
\cite{Kusmartsev:1989nc,Tamaki:2011zza}.
The next branch from $Q_{A}$ to $Q_{B}$ is unstable, 
and the branch from $Q_{B}$ to $Q_{C}$ is stable again
\cite{Tamaki:2011zza}.
This comprises a complete Whitney surface 
with a stable upper and lower sheet and an unstable area inbetween.
The next turning points are part of the spiral. 
Such spirals are described by Arnold \cite{Arnol'd:1992} 
as ``limit cycles'', where the equilibrium states lose their stability. 
Then all configurations, which are part of the spirals, are unstable.
We note, that this analysis is in complete agreement
with the discussion above for compact fermionic stars.

\begin{figure}[h!]
\begin{center}
\includegraphics[width=.45\textwidth, angle =0]{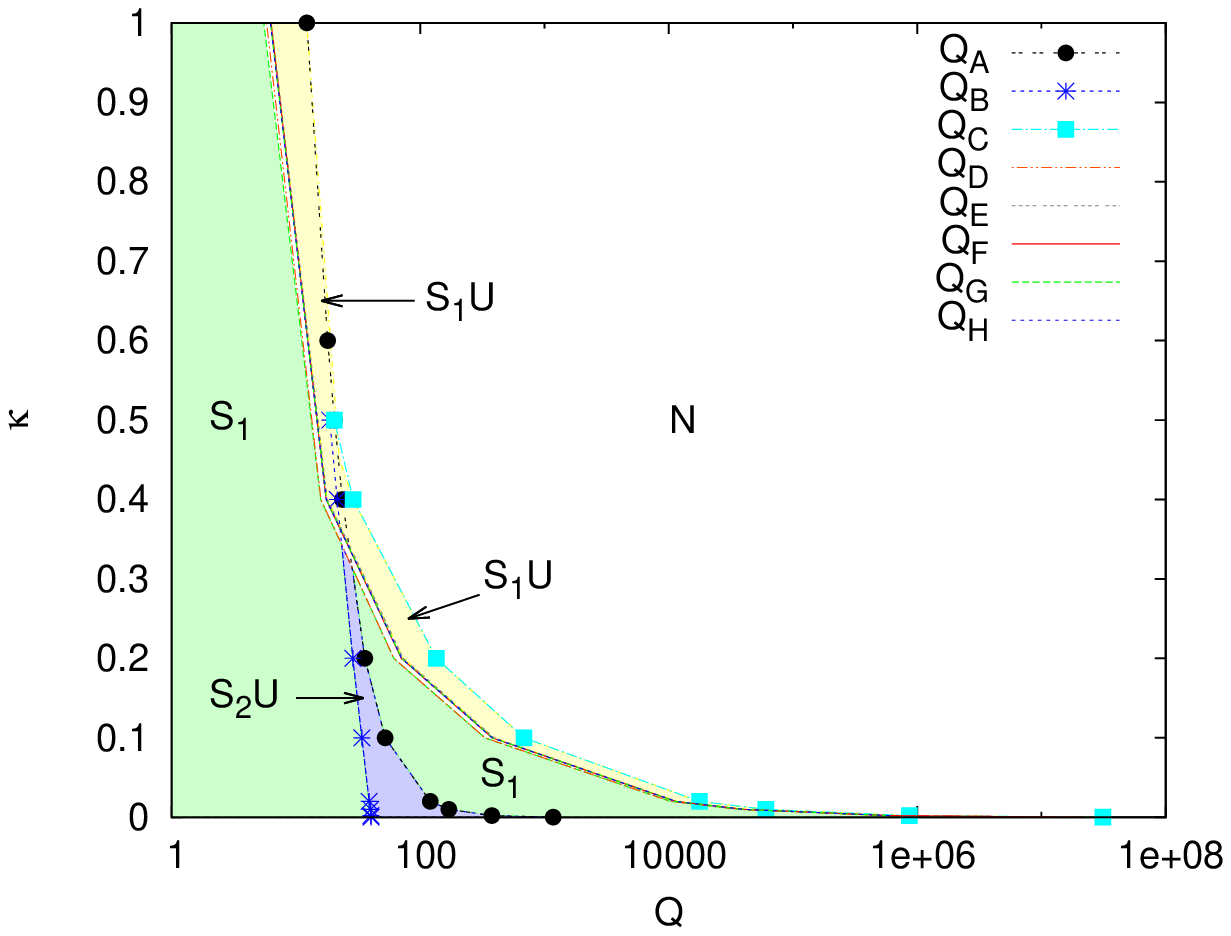}
\includegraphics[width=.45\textwidth, angle =0]{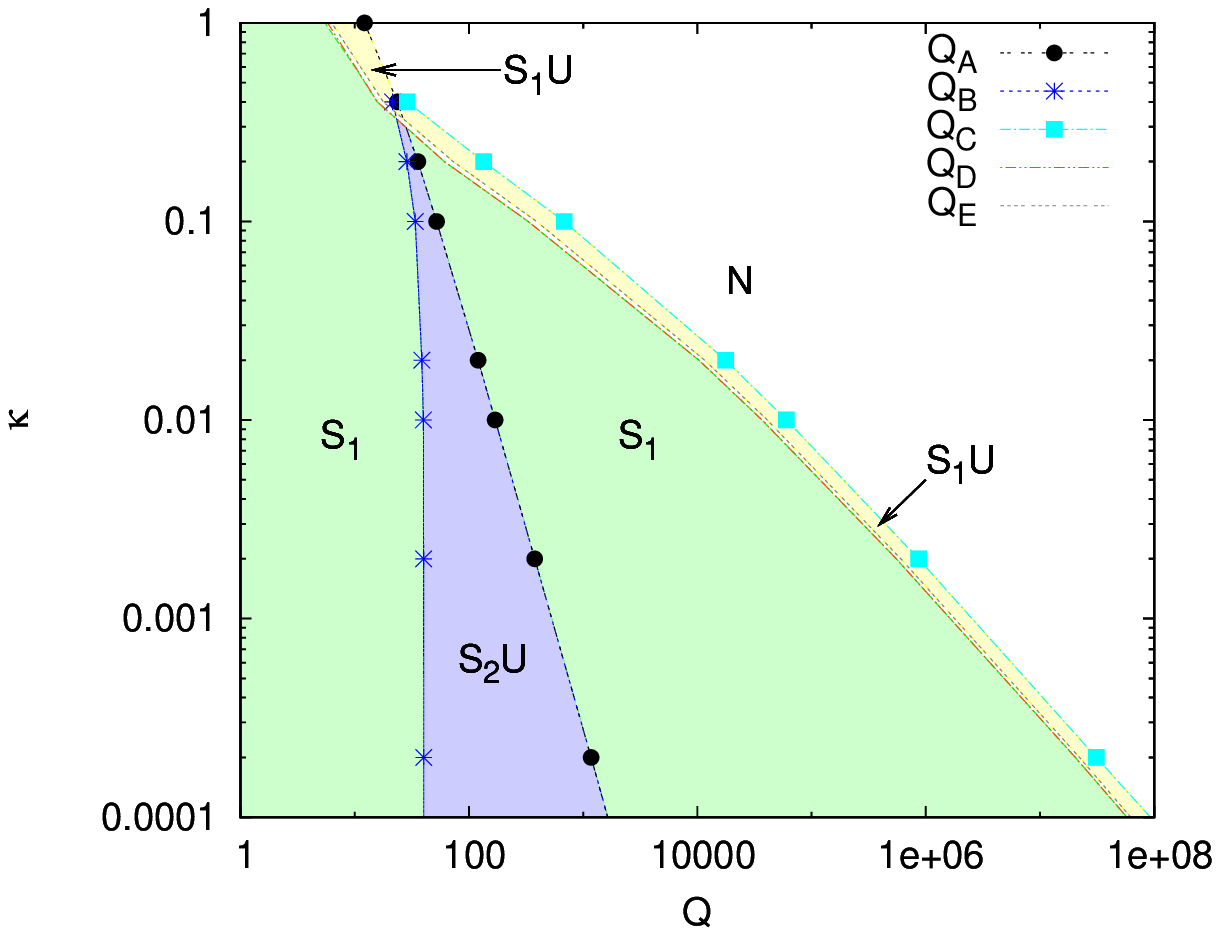}
\caption{\small Control space $\mathcal{C} = \{\kappa, Q \}$ 
for the fundamental boson stars. 
In the areas denoted by $S_1$, $S_i U ~ (i =1,2)$ and $N$, 
there is a single stable solution, there are $i$ stable solution(s) 
and one or more unstable solution(s), 
and there is no equilibrium solution, respectively.
The axes are semi-logarithmic (left) and double logarithmic (right).
}
\label{ContFundBS}
\end{center}
\end{figure}

Let us finally exhibit the stability properties of the boson stars by
presenting the control space $\mathcal{C} = \{\kappa, Q \}$
in Fig.~\ref{ContFundBS}.
The control space is the projection 
of the catastrophe map $\chi (\mathcal{M})$ 
into the control plane.
In the regions denoted by $S_1$, $S_i U ~ (i =1,2)$ and $N$, 
there is one stable solution, there are $i$ stable solution(s) 
and one or more unstable solution(s), 
and there are no equilibrium solution, respectively. 
The area delimited by the turning points $Q_{C}$ and $Q_{D}$ 
contains beside one stable solution all the unstable solutions
of the spiral.

The control space reveals that there are
two areas, where only a single stable solution exists.
The first area is associated with the stable branch terminating at $A$.
This is the branch formed by the stable lower density boson stars.
The second area is associated with the branch from $B$ to (almost) $C$.
This branch comprises the stable high density boson stars.
As the coupling constant $\kappa$ increases, 
the width of the second area diminishes and shrinks to zero
at the critical value of $\kappa$.
This transition is consistent with the fact that for large $\kappa$ 
the solutions tend to mini-boson star solutions, 
which possess only a single stable branch.

\subsection{Rotating Boson Stars}

Let us now turn to rotating boson stars.
These stationary axially symmetric configurations
have a finite angular momentum proportional to their particle number,
$J=n \, Q \ $ (Eq.~\ref{JnQ}),
with the ``quantum number'' $n \ne 0$. 
In the following we extend our above analysis
of the physical properties of boson stars to the rotating case,
focussing on solutions with $n=1$ and positive parity,
i.e.~$n = 1^+$ boson stars,
that were obtained previously \cite{Kleihaus:2005me,Kleihaus:2007vk}.

\subsubsection{Equilibrium space of $n = 1^+$ boson stars}

\begin{figure}[h!]
\begin{center}
\includegraphics[width=0.55\textwidth, angle =0]{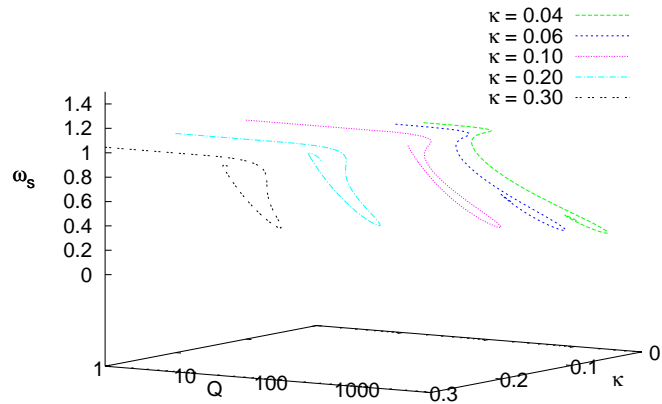}
\end{center}
\vspace{-0.5cm}
\caption{\small
{\it Equilibrium space} $\mathcal{M} = \{\omega_s, \kappa, Q \}$ 
for rotating boson stars with $n= 1^+$
in the range $0.04 \le \kappa \le 0.3$.
}
\label{Whitn1}
\end{figure}

As in the case of the spherically symmetric boson stars,
we begin our analysis of the $n = 1^+$ rotating
boson stars by considering 
the equilibrium space $\mathcal{M} = \{\omega_s, \kappa, Q \}$.
Again, for a fixed value of the dimensionless coupling constant $\kappa$
a family of stationary solutions exists in the frequency range
$\omega_0(\kappa) \le \omega_s \le \omega_{\rm max}$,
where $\omega_0(\kappa)$ increases with $\kappa$,
whereas $\omega_{\rm max}$ is still given by the boson mass
\cite{Kleihaus:2005me}.
This is seen in Fig.~\ref{Whitn1}, where
we illustrate the equilibrium space $\mathcal{M} = \{\omega_s, \kappa, Q \}$
of the $n = 1^+$ boson stars.

Clearly, the structure of the
equilibrium space $\mathcal{M} = \{\omega_s, \kappa, Q \}$
of the $n=1^+$ boson stars is very similar
to the one of the $n=0$ solutions.
All families of solutions (for fixed $\kappa$)
start from the vacuum $Q=0$ at the upper limit
of the frequency, and all form spirals at the lower end.
However, we note that as $\kappa$ increases, 
the $n=1^+$ spirals become more elongated in this
equilibrium space with respect to $\omega_s$
than their $n=0$ counterparts.

We could also consider the set of equilibrium solutions
with respect to other sets of parameters.
For the $n=0$ boson stars we considered
the alternative equilibrium space $ \mathcal{N} = \{ \phi_0, M, \kappa \} $.
Here we would have to replace the variable $\phi_0$
by another variable, however, since for rotating
boson stars $\phi_0=0$.
Such an alternative variable for $n=1^+$ boson stars would be $\phi_0'=\phi'(0)$
\cite{Hartmann:2010pm}.

\subsubsection{Cusp structure of $n=1^+$ boson stars}

Let us next turn to the cusp structure of these
rotating boson stars.
As in the case of the $n=0$ solutions,
for not too large values of the coupling constant $\kappa$
a first branch of rotating boson star solutions
resides between the vacuum $M=Q=0$  
and the local maximum of the mass and charge,
$M_A$, $Q_A$, at the first cusp $A$.
Next, a second branch connects the first cusp $A$ with the second cusp $B$,
where the mass and charge have a local minimum,
$M_B$, $Q_B$.
Then a third branch emerges from the second cusp $B$
and extends up to the third cusp $C$,
where the mass and charge reach their global maximum,
$M_C$, $Q_C$.
Beyond $C$, finally, a spiral is formed.
We list these turning points in Tab.~\ref{tabQNn1}. 

\begin{table}[h!]
\begin{center}
\begin{tabular}{|c|c|c|c|c|c|c|}
\hline
$\kappa$ & $Q_{A}$ & $Q_{B}$ & $Q_{C}$ & $Q_{D}$    \\
\hline
\hline
$0.04$ & $226.888$ & $135.570$ & $5121.20$ & $1630.19$  \\
$0.06$ & $184.8$   & $130.123$ & $2408.91$ & $434.076$  \\
$0.10$ & $140.631$ & $118.113$ & $910.699$ & $77.434$   \\
$0.20$ & $96.197$  & $94.1$    & $237.173$ & $35.984$   \\
\hline
$0.30$ & $-$       & $-$       & $110.958$ & $23.462$   \\
 \hline
\end{tabular}
\end{center}
\caption{Turning points for rotating boson stars with $n = 1^+$ 
for several values of the
coupling constant $\kappa$. 
}
\label{tabQNn1}
\end{table}

\begin{figure}[h!]
\begin{center}
\includegraphics[width=0.4\textwidth, angle =0]{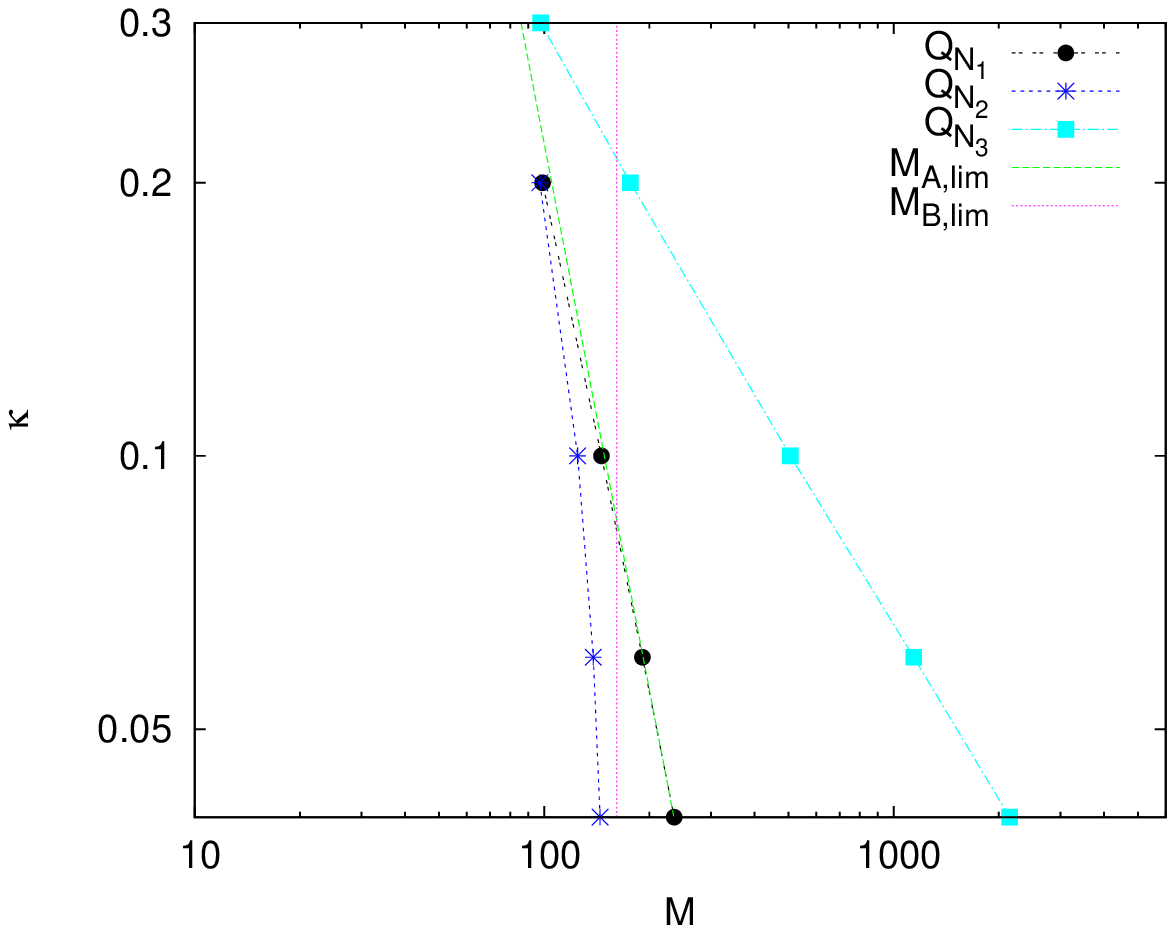}
\includegraphics[width=0.4\textwidth, angle =0]{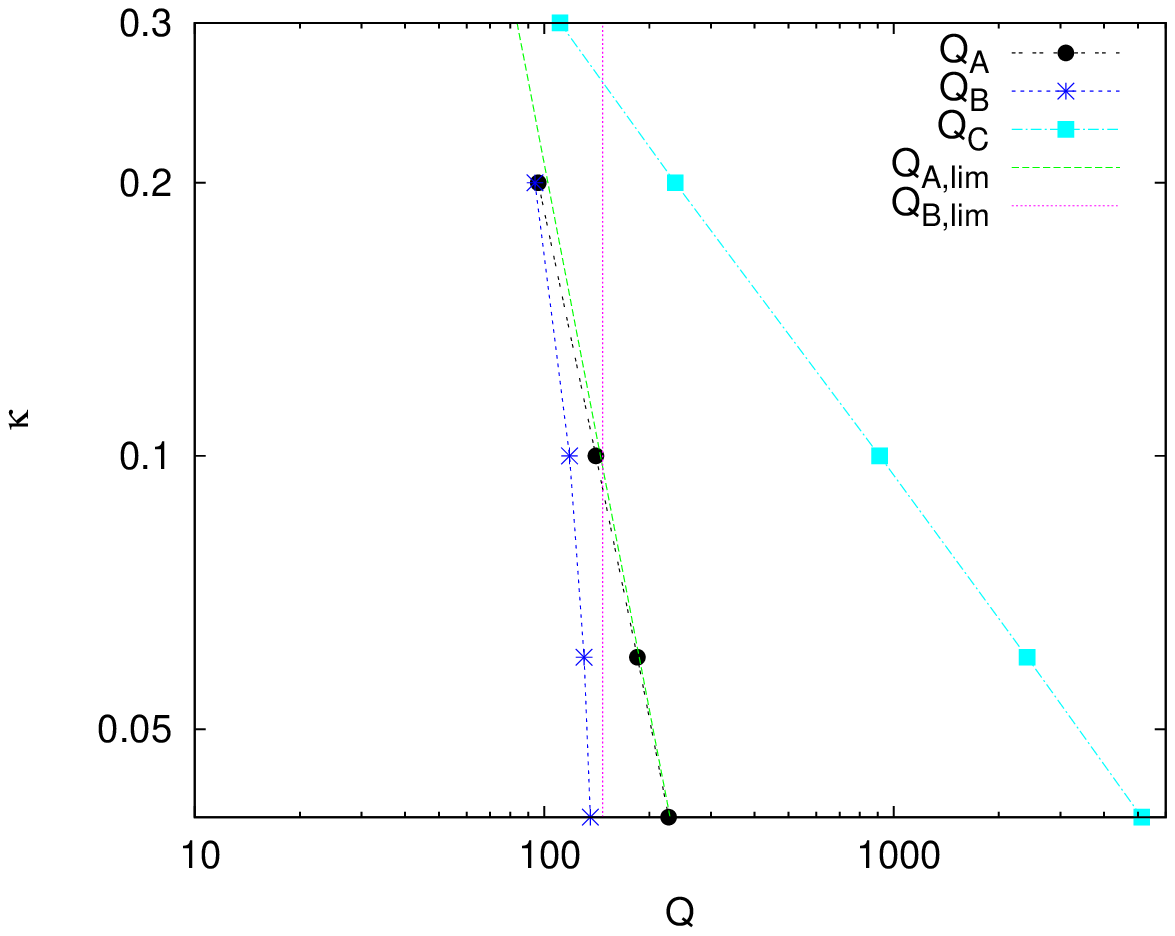}
\caption{\small
Coupling constant $\kappa$
versus the mass $M$ (left) and
versus the particle number $Q$ (right)
for the cusps $A$, $B$ and $C$,
together with functions approximating the behavior
of $A$ and $B$ in the limit $\kappa \to 0$.
Note, that for the larger values of $\kappa$
the cusps $A$ and $B$ are no longer present.}
\label{cusp-rot}
\end{center}
\end{figure}

Analogous to the non-rotating case,
the values of the mass $M_A$ and the charge $Q_A$ at
the cusp $A$ increase,
as $\kappa \to 0$. 
Indeed, we observe the same $\kappa^{-1/2}$ dependence,
as seen in Fig.~\ref{cusp-rot}.
The values of the mass $M_B$ and the charge $Q_B$ at the cusp $B$,
on the other hand, are expected to tend to constant values
in the limit $\kappa \to 0$, corresponding now to the
values of the mass and the charge of the unique minimum
of the rotating $Q$-ball solutions in flat space-time.
(The available data do not yet suffice to fully demonstrate this behaviour, 
however.)
The values of the mass $M_C$ and the charge $Q_C$ at the cusp $C$
increase again without bound as $\kappa \to 0$.
The available data do not yet exhibit
the expected limiting $\kappa^{-3/2}$ dependence,
but a somewhat deviating $\kappa$ dependence.
This is analogous to the non-rotating case in this range of values of
the coupling constant $\kappa$, where 
the limiting $\kappa^{-3/2}$ dependence is only approached
for considerably smaller values of $\kappa$.

For larger values of $\kappa$, however, we observe a somewhat different
pattern for the cusps $A$, $B$ and $C$ 
in the case of the rotating boson stars as compared to the non-rotating case.
In the rotating case, the extrema $A$ and $B$ merge, and disappear
at a critical value of $\kappa$, while $C$ remains,
as seen in Tab.~\ref{tabQNn1} and Fig.~\ref{cusp-rot}.
In the non-rotating case it is $B$ and $C$, which merge,
while $A$ remains.
In both cases, however, for large values of $\kappa$
the solutions exhibit the same pattern as the
mini-boson stars: they possess a single physically relevant branch,
beyond which a spiral is formed.

\subsubsection{Size of $n=1^+$ boson stars}

\begin{figure}[h!]
\begin{center}
\includegraphics[width=.45\textwidth, angle =0]{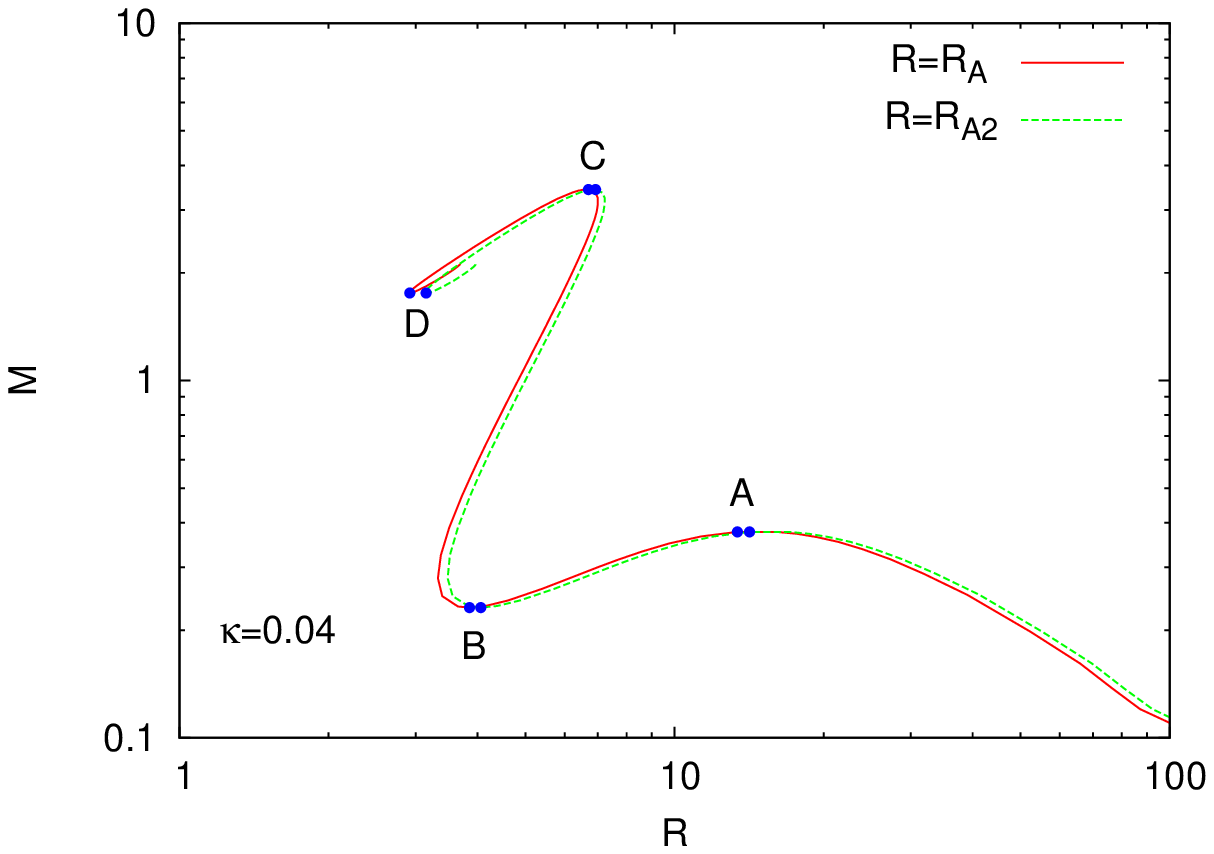}
\includegraphics[width=.45\textwidth, angle =0]{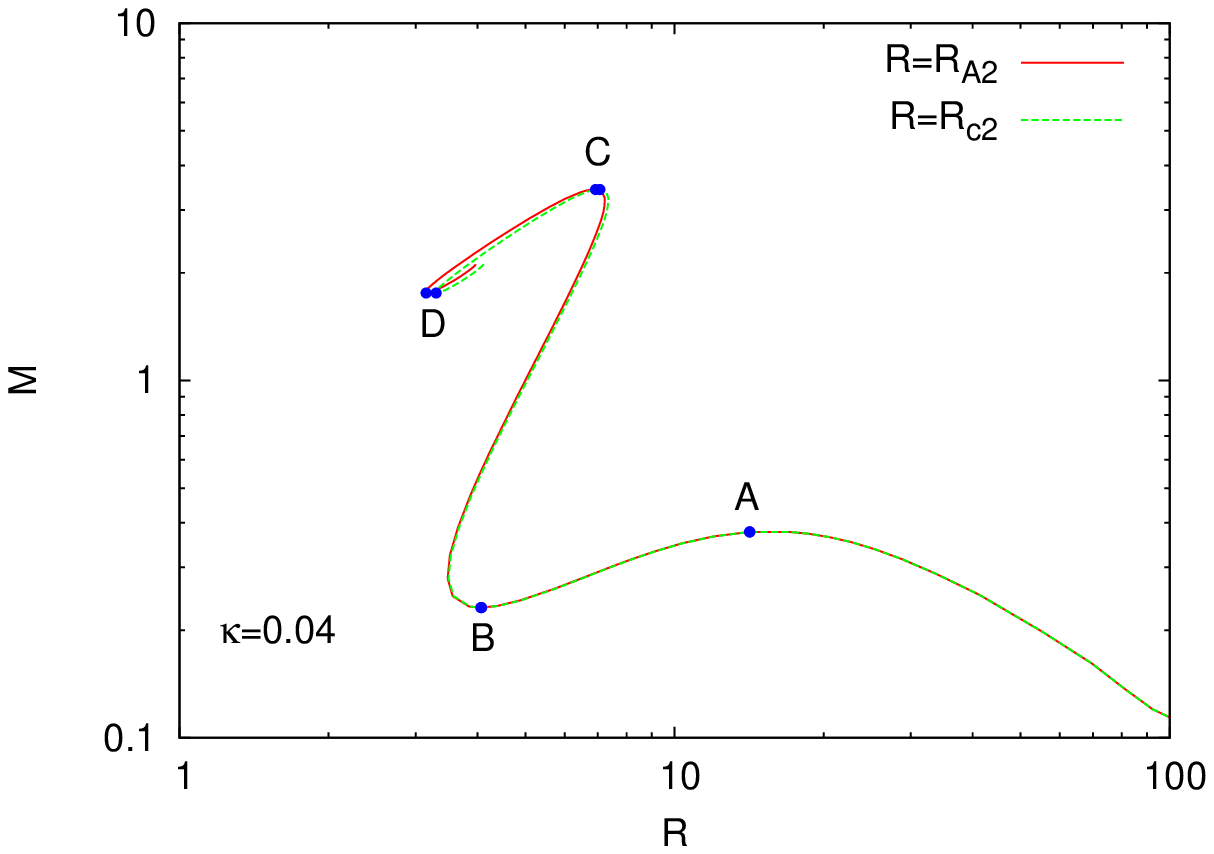}
\caption{\small A comparison of different radius definitions for
$n=1^+$ rotating boson stars:
$R_A$ and $R_{A2}$ (left),
$R_{A2}$ and $R_{C2}$ (right)
for $\kappa=0.04$.
}
\label{radii_rot}
\end{center}
\end{figure}

As in the case of the $n=0$ boson stars,
we would like to study the compactness of the
rotating boson stars
and their proximity to the black hole limit.
For that purpose we consider several
possiblities to obtain a measure for the size 
of these axially symmetric boson star solutions.
We define area type radii via introducing first 
the function $R(r)$ 
\begin{equation}
R(r) = \left( \frac{1}{4 \pi} \int_\Omega \left.
\sqrt{ g_{\varphi\varphi} g_{\theta\theta} }\right|_r \, d\Omega
\right)^{1/2} \ , 
\label{Rr1}
\end{equation}
and then using this $R(r)$ to obtain as measures for the size
\begin{equation}
R_A= \frac{ \int j^t R(r) \sqrt{-g} \, d^3 r}{Q}
\label{RA}
\end{equation}
and
\begin{equation}
R_{A2}^2 = \frac{ \int j^t (R(r))^2 \sqrt{-g} \, d^3 r}{Q} \ .
\label{RA2}
\end{equation}
Similarly, we define circumferential type radii
via first introducing the function
\begin{equation}
R(r) = \frac{1}{2 \pi} \int_C \left. \sqrt{ g_{\varphi\varphi} }
\right|_{r,\theta=\pi/2} \, d \varphi
\label{Rr2}
\end{equation}
and then using this $R(r)$ to obtain $R_C$ as in Eq.~(\ref{RA})
and $R_{C2}^2$ as in Eq.~(\ref{RA2}).
All these definitions give rather similar results.
We demonstrate this (in part) in Fig.~\ref{radii_rot},
where we compare the radii $R_{A2}$, $R_A$  and $R_{C2}$.

The figure clearly shows, that
the dependence of the mass on the size is very similar
for rotating and non-rotating boson stars.
Focussing on the smaller values of $\kappa$
(away from the mini-boson star limit) 
the mass increases along the first branch,
while the size decreases,
until the cusp $A$ is reached,
marking the end of the lower density phase.
Between $A$ and $B$ the mass decreases with decreasing size.
Beyond $B$ the mass then rises steeply again,
while the size also increases along (most of) this third branch.
This high density phase
of the rotating boson stars then ends at the cusp $C$,
where the spiral starts.\\

\subsubsection{Black hole limit}

\begin{figure}[h!]
\begin{center}
\includegraphics[width=.45\textwidth, angle =0]{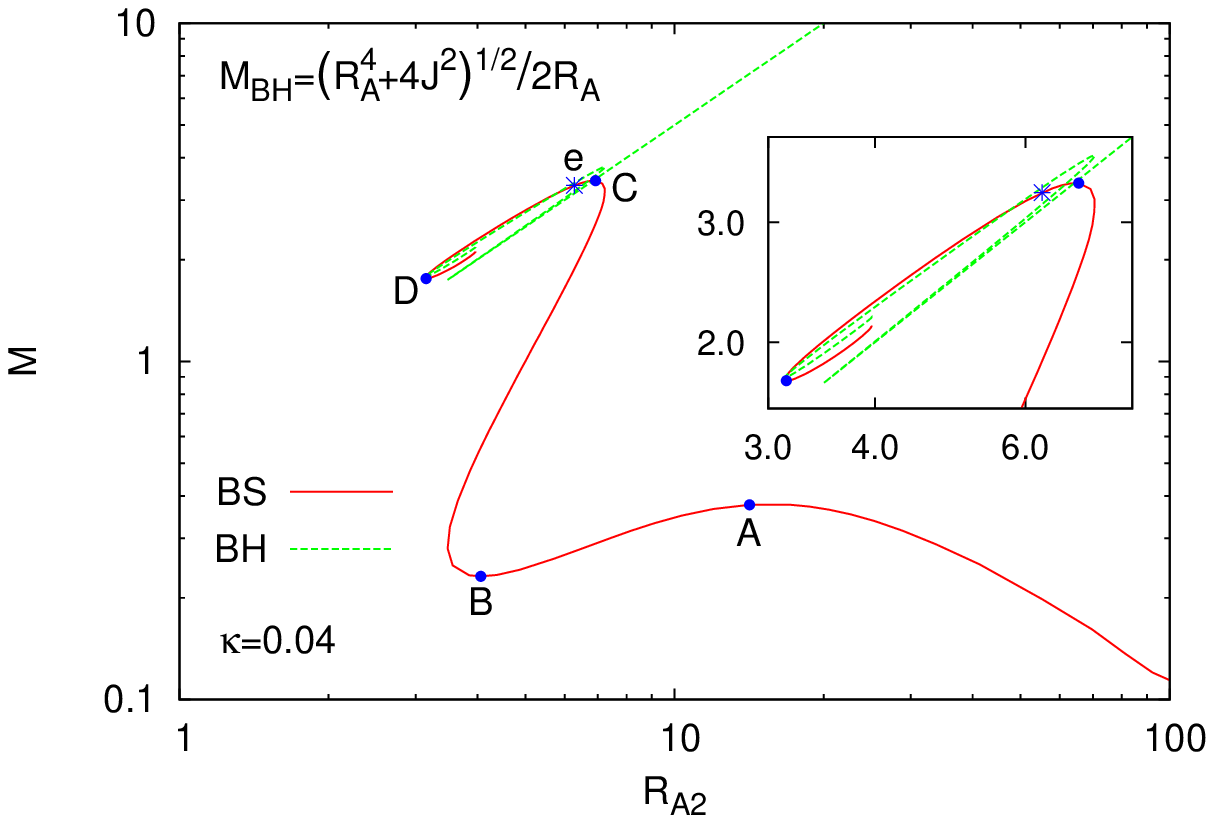}
\includegraphics[width=.45\textwidth, angle =0]{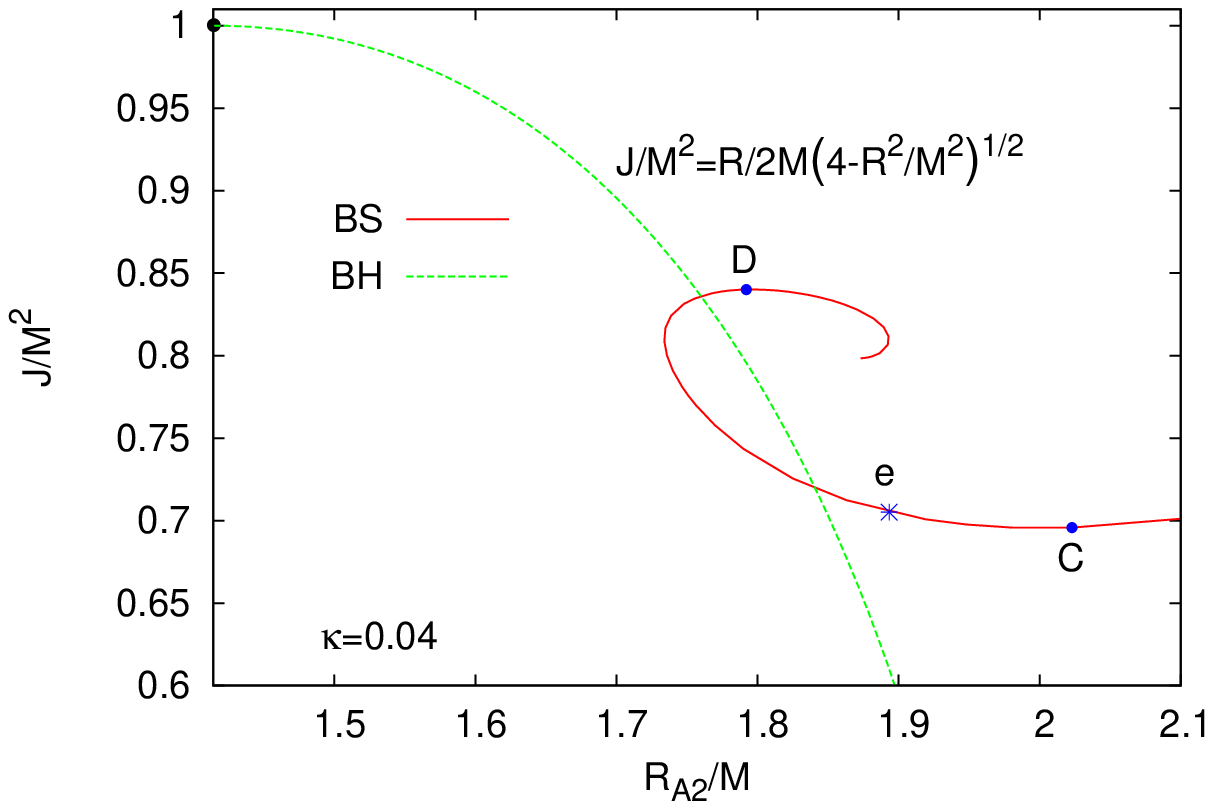}
\caption{\small Comparison of the boson star mass $M$ versus the size $R_{A2}$
for $\kappa=0.04$
with the mass $M_{\rm BH}$ of the Kerr black hole
obtained for the same size $R$ and angular momentum $J$
(left);
comparison of the ratio $J/M^2$ versus the ratio $R_{A2}/M$ 
of the boson stars 
($\kappa=0.04$) with the Kerr black hole curve (right).
The point $e$ indicates the onset of an ergoregion.
}
\label{radii_bh}
\end{center}
\end{figure}

As in the case of the non-rotating boson stars,
we would now like to
address the black hole limit for these rotating boson stars.
For this purpose we employ these radii to compare the boson star
masses and sizes with those of the corresponding rotating Kerr black holes.
In particular, for the set of values of the size $R_{A2}$ 
and angular momentum $J$ of the boson stars,
their mass $M$ is compared with 
the mass $M_{\rm BH}$ of the corresponding Kerr black holes,
which are evaluated for the same values of $R_{A2}$ and $J$,
where $R_{A2}$ is defined via the event horizon area.
We exhibit this comparison in Fig.~\ref{radii_bh} for $\kappa=0.04$.
We note, that because of this construction of the Kerr curve,
it has as many branches as the boson star curve.
Interestingly, 
we observe that as in the case $n=0$, also the rotating
boson stars are very close to the
black hole limit, when the spiral is formed,
as seen in Fig.~\ref{radii_bh} (left).
(The spiral is enlarged in the inset, which also features
the Kerr black hole branch corresponding to the steep high density
boson star branch.)

For a given mass $M_{\rm BH}$, the angular momentum $J_{\rm BH}$
of a Kerr black hole cannot exceed a certain bound,
the Kerr bound,  $J_{\rm BH}/M_{\rm BH}^2 \le 1$.
The Kerr bound is saturated for extremal black holes.
Higher angular momenta correspond to naked singularities.
When considering this scaled angular momentum $J_{\rm BH}/M_{\rm BH}^2$
versus the scaled area $A_{\rm BH}/M_{\rm BH}^2$ 
of the black holes,
all Kerr black holes fall onto a single line,
starting from the point corresponding to the
set of Schwarzschild solutions and ending
at the point corresponding to the set of extremal Kerr solutions.
The upper part of this line 
representing fast rotating
Kerr black holes is exhibited in Fig.~\ref{radii_bh} (right).

In our comparison of the rotating boson stars with the Kerr black holes
we would now like to know, how close the boson stars
are to this Kerr bound, when they are highly compact.
Therefore we also show
$J/M^2$ versus $R_{A2}/M$ for such a set of highly compact
rotating boson stars in Fig.~\ref{radii_bh} (right).
In particular, we have also indicated the point $C$ of the boson star curve, 
where the spiral starts, and the point $D$, 
which is located inside the spiral.
We note that the family of boson stars 
assumes at $C$ its minimum value of $J/M^2$.
For $\kappa=0.04$, the case shown in the figure,
$J/M^2\approx 0.7$ at $C$.
When $\kappa$ increases, the value of $J/M^2$ at $C$ increases
($\kappa=0.06$: 0.78, $\kappa=0.1$: 0.86, 
$\kappa=0.2$: 0.92, $\kappa=0.3$: 0.94).
Thus these highly compact boson stars
close to the black hole limit
can rotate with angular momenta close to the Kerr bound.

In Fig.~\ref{radii_bh} (right)
we have also indicated the point, beyond which
the rotating boson stars possess ergoregions
\cite{Kleihaus:2007vk}.
For $\kappa=0.04$, the formation of ergoregions
arises only inside the spiral.
For larger values of $\kappa$ 
the ergoregion formation starts already on
the high density branch shortly before $C$ is reached.
We therefore conclude that for these $n=1^+$ rotating boson stars
there appears to be a correlation between
approaching the black hole limit and developing
an ergoregion.\\

\subsubsection{Stability analysis of $n=1^+$ boson stars}

\begin{figure}[h!]
\begin{center}
\includegraphics[width=0.45\textwidth, angle =0]{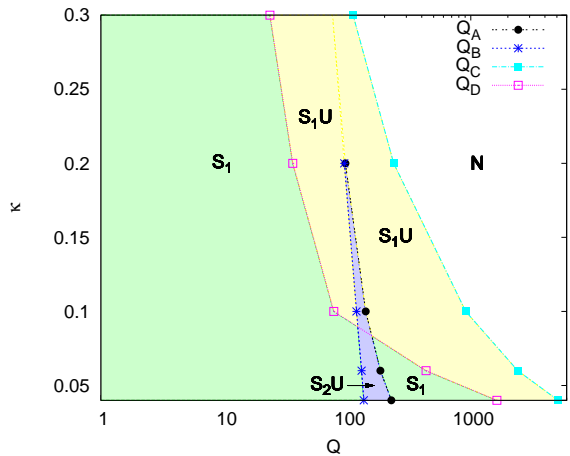}
\end{center}
\vspace{-0.5cm}
\caption{\small
{\it Control space} $\mathcal{C} = \{ \kappa, Q \}$
for rotating boson stars with $n= 1^+$ in the range $0.04 \le \kappa \le 0.3$.
}
\label{Whitn2}
\end{figure}

To address the stability of these rotating boson stars, we turn to 
the control space $\mathcal{C} = \{ \kappa, Q \}$,
presented in Fig.~\ref{Whitn2}.
Since both $n=0$ and $n=1^+$ boson stars have analogous
{\it equilibrium spaces} $\mathcal{M} = \{\omega_s, \kappa, Q \}$,
the corresponding discussion of their stability
is also analogous.
From the analysis via catastrophe theory
we conclude that there are 
two areas which possess only a single stable solution.
The first area is associated with the stable branch terminating at $A$,
formed by the stable lower density boson stars.
The second area is associated with the branch from $B$ to (almost) $C$,
comprising the stable high density boson stars.
As the coupling constant $\kappa$ increases,
the width of the second area diminishes and shrinks to zero
at a critical value of $\kappa$.
The width of the area labeled to represent 2 stable solutions
as well as unstable solutions also diminishes with increasing $\kappa$.
The width shrinks to zero at the critical value of $\kappa$, 
where the turning points $A$ and $B$ merge
and disappear, while only $C$ remains (apart from the 
critical points of the spiral).

According to catastrophe theory arguments
the branch from the vacuum 
to the first turning point $Q_{A}$ is a stable one.
The next branch from $Q_{A}$ to $Q_{B}$ is unstable,
while the branch from $Q_{B}$ to $Q_{C}$ should be stable again.
However, we must also take into account the emergence of ergoregions,
which are associated with their own kind of instability
\cite{Friedman:1978,Comins:1978,Yoshida:1996,Cardoso:2007az}.
While for small values of $\kappa$ ergoregions arise only
beyond $C$ somewhere in the spiral,
which is unstable anyway,
for larger values of $\kappa$ ($\kappa \ge 0.1$)
ergoregions appear already
along the higher density branch from $B$ to $C$,
signalling an instability of these highly compact rotating
boson stars close to $C$
\cite{Kleihaus:2007vk,Cardoso:2007az}.

Apart from this new type of instability,
present only in the rotating case,
the boson star solutions for the non-rotating ($n=0$)
and the rotating ($n=1^+$) case exhibit a similar
general pattern.

\section{Conclusions}

We have addressed the physical properties 
of non-rotating and rotating boson stars,
obtained with a self-interaction potential of the scalar field,
which allows for non-topological soliton solutions
in the absence of gravity.
Such a self-interaction potential is crucial
to find a much richer set of solutions
than the ones obtained with only a mass term (mini-boson stars)
and with a repulsive $|\Phi|^4$ interaction.

In particular, we note that there are two stable
regions in the equilibrium space of the boson star solutions,
when this solitonic self-interaction potential is employed,
whereas there is only one such stable region for boson stars
which do not possess a flat space-time limit.
In fact, there is an interesting analogy with compact stars,
which possess a lower density phase, the white dwarf phase,
and a high density phase, the neutron star phase,
since the boson stars also exhibit a lower density phase
and a high density phase.
Moreover, beyond the neutron star phase the compact stars
exhibit an unstable spiralling phase very close to the
black hole limit. Such an unstable spiralling phase very close
to the black hole limit is also seen for the boson stars
beyond their stable high density phase.
 
A stable mini-boson star is an equilibrium state,
where the Heisenberg uncertainty principle $\Delta r \Delta p \sim \pi \hbar$ 
provides the means to balance gravity and avoid collaps,
below a (small) critical mass of the mini-boson stars \cite{Mielke:2000mh}.
In boson stars with a repulsive $|\Phi|^4$ potential term
this self-interaction allows for much larger stable boson stars
\cite{Colpi:1986ye}.
The solitonic self-interaction potential, on the other hand,
has repulsive and attractive components, which 
dominate the features of the solutions in a large region
of the equilibrium space,
leaving here for gravity only a minor role to play.

Indeed, 
for not too large values of the coupling constant $\kappa$,
the properties of the solutions follow rather closely those
of the corresponding non-topological solitons.
Only at the boundaries of the domain of existence,
gravity becomes dominating. Here in flat space-time the solutions
would grow without limit. 
Thus the single infinitely long stable branch
of the non-topological solitons
is reflected in the finite (mostly) stable branch $B-C$ of the boson stars.
Along (most of) this branch the radius increases as the mass increases.
This branch ends, 
when the boson star's compactness approaches the black hole limit.
The soliton solutions would simply cross this limit. 
But as long as the coupling to gravity is finite,
no matter how small it is, this limit cannot be exceeded.
Thus collapse is unavoidable, 
as signalled by the formation of a spiral in the
equilibrium space of these stationary solutions.

The single infinitely long unstable branch
of the non-topological solitons, on the other hand,
is reflected in the finite unstable branch $A-B$ of the boson stars.
But gravity allows for an additional stable branch $0-A$,
present even in the case of no self-interaction.
Here the size of the solutions decreases as the mass increases, 
which reveals the dominating influence of gravity
for this branch.

When comparing the non-rotating ($n=0$) and the rotating ($n=1^+$) boson stars
we note, that many of their features are very similar.
While the high density soliton-type boson star branch 
is bounded by the Schwarzschild black holes in the $n=0$ case, 
it is bounded by the Kerr black holes in the $n=1^+$ case.
Because the centrifugal force will counteract the gravitational force,
it is expected that rotation stabilizes a boson star \cite{Mielke:2000mh},
and the available data indeed show, that in the presence of rotation
for given values of $\kappa$ higher masses are reached
for the rotating boson stars.
However, rotation also comes with a {\sl caveat} for stability,
since for globally regular objects such as boson stars
the presence of an ergoregion implies an instability,
associated with superradiant scattering
\cite{Friedman:1978,Comins:1978,Yoshida:1996,Cardoso:2007az}.

Thus rotating boson stars become unstable, 
when they develop an ergoregion.
For the $n=1^+$ boson stars considered,
the ergoregion is formed either inside the spiral
or very close to the cusp $C$.
Therefore the possible presence of ergoregions 
put forward by Cardoso et al.~\cite{Cardoso:2007az} as an argument 
to exclude boson stars and various other black hole doubles
as potential horizonless candidates for compact dark astrophysical objects,
is not yet compelling.
It remains to be seen, whether boson stars with
appropriate values of the physical parameters
to fit observational data
will or will not suffer from such an ergoregion instability.
Our analysis shows, that there are stable highly compact
boson stars close to the black hole limit,
that are at the same time close to the Kerr bound $J/M^2=1$.

\begin{figure}[h!]
\begin{center}
\includegraphics[width=0.55\textwidth, angle =0]{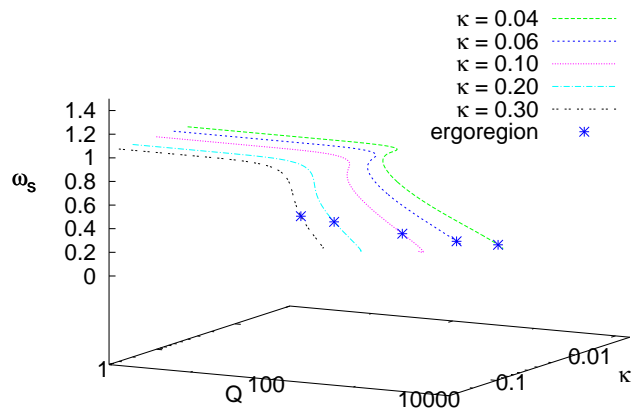}
\end{center}
\vspace{-0.5cm}
\caption{\small
{\it Equilibrium space} $\mathcal{M} = \{\omega_s, \kappa, Q \}$
for rotating boson stars with $n= 2^+$ in the range $0.04 \le \kappa \le 0.3$.
The onset of the ergoregion is indicated by the asterisks.
}
\label{WhitN2Plus}
\end{figure}

Finally, we would like to address the case of rotating boson stars with higher
rotation quantum numbers $n\ge 2$. Such boson stars have been addressed
by Ryan \cite{Ryan:1996nk}.
Since the numerical analysis for these systems is difficult
\cite{Kleihaus:2005me,Kleihaus:2007vk},
we have not yet accumulated a sufficient set of solutions
to give a similar discussion of their properties
as for the $n=0$ and $n=1^+$ boson stars.
Still, we present in Fig.~\ref{WhitN2Plus} a part of 
the equilibrium space $\mathcal{M} = \{\omega_s, \kappa, Q \}$
of the $n = 2^+$ boson stars.
Whereas still higher values of the mass are reached on the
soliton-type branch, we note, that the onset of the ergoregion
instability happens earlier, thus decreasing the viablity range
of these solutions.

Clearly, further calculations of fast rotating boson stars are called for,
not only in the solitonic model but also in $|\Phi|^4$-type models,
where a possible ergoregion instability was not yet considered.
Most interesting will, however, be an extension of the investigation of the 
dynamical evolution of boson stars \cite{Seidel:1990jh,Balakrishna:1997ej}
to the rotating case.

\section*{Acknowledgements}
We would like to thank Meike List and Isabell Schaffer for discussions
and in particular for sharing their data with us.
B.K.~gratefully acknowledges support by the DFG.

\appendix

\section{Construction of the solutions}

\subsection{Boundary conditions}

The choice of appropriate boundary conditions must guarantee 
that the boson star solutions
are globally regular and asymptotically flat,
and that they possesses finite energy and finite energy density.

For spherically symmetric boson stars
boundary conditions must be specified
for the metric functions $f(r)$ and $l(r)$ 
and the scalar field function $\phi(r)$ 
at the origin and at infinity.
At the origin one finds the boundary conditions
\begin{equation}
\partial_r f|_{r=0}=0 \ , \ \ \  
\partial_r l|_{r=0}=0 \ , \ \ \
\partial_r \phi|_{r=0}=0 \ .
\label{bc1} \end{equation}
Note, that for spherically symmetric boson stars the scalar field
has a finite value $\phi_0$ at the origin,
\begin{equation}
\phi(r)= \phi_0 \, + O(r^2 ) \ . \label{phi_r0}
\end{equation}
For $r \rightarrow \infty$ 
the metric approaches the Minkowski metric $\eta_{\alpha \beta}$
and the scalar field assumes its vacuum value $\Phi=0$. 
Accordingly, we impose at infinity the boundary conditions 
\begin{equation}
f|_{r \rightarrow \infty}=1 \ , \ \ \
l|_{r \rightarrow \infty}=1 \ , \ \ \
\phi|_{\, r \rightarrow \infty}=0 \ . 
\label{bc2} \end{equation}

For rotating axially symmetric boson stars
appropriate boundary conditions must be specified
for the metric functions $f(r,\theta)$, $l(r,\theta)$, $h(r,\theta)$
and $\omega(r,\theta)$, 
and the scalar field function $\phi(r,\theta)$
at the origin, at infinity, on the positive $z$-axis ($\theta=0$),
and, exploiting the reflection symmetry w.r.t.~$\theta \rightarrow
\pi - \theta$, in the $xy$-plane ($\theta=\pi/2$).
At the origin we require
\begin{equation}
\partial_r f|_{r=0}=0 \ , \ \ \
\partial_r l|_{r=0}=0 \ , \ \ \
h|_{r=0}=1 \ , \ \ \
\omega|_{r=0}=0 \ , \ \ \
\phi| _{r =0}=0 \ .
\label{bc3} \end{equation}
At infinity the boundary conditions are
\begin{equation}
f|_{r \rightarrow \infty} =1 \ , \ \ \
l|_{r \rightarrow \infty} =1 \ , \ \ \
h|_{r \rightarrow \infty} =1 \ , \ \ \
\omega|_{r \rightarrow \infty} =0 \ , \ \ \
\phi| _{r \rightarrow \infty}=0 \ ,
\label{bc4} \end{equation}
and for $\theta=0$ and $\theta=\pi/2$, respectively, 
we require the boundary conditions
\begin{equation}
\partial_{\theta} f|_{\theta=0}=0 \ , \ \ \
\partial_{\theta} l|_{\theta=0}=0 \ , \ \ \
h|_{\theta=0}=1 \ , \ \ \
\partial_{\theta} \omega |_{\theta=0}=0 \ , \ \ \
\phi |_{\theta=0}=0 \ , 
\label{bc5} \end{equation}
and for even parity solutions
\begin{equation}
\partial_{\theta} f|_{\theta=\pi/2}=0 \ , \ \ \
\partial_{\theta} l|_{\theta=\pi/2}=0 \ , \ \ \
\partial_{\theta} h|_{\theta=\pi/2}=0 \ , \ \ \
\partial_{\theta} \omega |_{\theta=\pi/2}=0 \ , \ \ \
\partial_{\theta} \phi |_{\theta=\pi/2}=0 \ ,
\label{bc6} \end{equation}
while for odd parity solutions
$\phi |_{\theta=\pi/2}=0$.

\subsection{Numerical methods}

First of all,
because of the power law fall-off of the metric functions,
we compactify space by introducing the compactified radial coordinate
\begin{equation}
\bar r = \frac{r}{1+r} \ .
\label{rcomp} \end{equation}
Then the resulting set of equations is solved numerically
subject to the above boundary conditions.

For spherically symmetric non-rotating solutions ($n=0$)
the set of equations depends only on the radial coordinate.
It is solved numerically
by employing a collocation method for boundary-value ordinary
differential equations developed by Ascher, Christiansen and Russell
\cite{COLSYS}.
Here the damped Newton method of quasi-linearization 
is applied. At each iteration step a linearized problem
is solved by using a spline collocation at Gaussian points.

Rotating solutions are obtained when $n \ne 0$.
The resulting set of coupled non-linear partial differential equations
is solved numerically 
by employing a finite difference solver \cite{FIDISOL},
based on the Newton-Raphson method.
The equations are discretized on a non-equidistant grid in
$\bar r$ and  $\theta$.
Typical grids used have sizes $90 \times 70$,
covering the integration region
$0\leq \bar r\leq 1$ and $0\leq\theta\leq\pi/2$.


\end{document}